%% file: nEXO_radio_assay.tex
\documentclass[%
 reprint,
 superscriptaddress,
 altaffilletter,
amsmath,amssymb,
aps,
prc,
floatfix,
]{revtex4-2}
\usepackage{graphicx}
\usepackage[caption=false]{subfig}
\usepackage{longtable}
\usepackage{xcolor}
\usepackage{tabularray}
\UseTblrLibrary{siunitx}
\usepackage{upgreek}
\usepackage{calc}
\usepackage[
colorlinks=true,
linkcolor=blue,
urlcolor=blue,
citecolor=blue
]{hyperref}
\usepackage{cleveref}
\crefformat{sample}{#2sample~#1#3}
\crefrangeformat{sample}{samples~#3#1#4 to~#5#2#6}
\crefmultiformat{sample}{samples~#2#1#3}  { and~#2#1#3}{, #2#1#3}{, and~#2#1#3}
\crefformat{figure}    {#2Fig.~#1#3}
\crefformat{table}     {#2Tab.~#1#3}
\crefname{sampleno}{sample}{sample}

\usepackage{pgf}
\DeclareFontFamily{U}{mathb}{}
\DeclareFontShape{U}{mathb}{m}{n}{<-5.5> mathb5 <5.5-6.5> mathb6 
<6.5-7.5> mathb7 <7.5-8.5> mathb8 <8.5-9.5> mathb9 <9.5-11> mathb10 
<11-> mathb12}{}
\DeclareSymbolFont{mathb}{U}{mathb}{m}{n}
\DeclareFontSubstitution{U}{mathb}{m}{n}
\DeclareMathSymbol{\blackdiamond}{\mathbin}{mathb}{"0C}

\input{units}

\input{isotopes}

\input{newcommands}

\usepackage{orcidlink}

\begin{document}
  \preprint{arXiv:26......}%
  \title{The nEXO Radioassay Program}% 

\input{authors-nexo-radioassay2}\date{\today}%
\begin{abstract}%
Material radioactivity compilations, such as the one presented here, are important enablers of science. They are useful for the selection of radiopure materials used in the design and construction of low-energy rare-event search experiments. They allow researchers developing such experiments to save time on material studies and avoid costly duplication of effort. The data presented here were generated in support of the planned nEXO double-beta decay search. This work contains among the most restrictive constraints on the natural radioactivity content of materials of general interest to the low-radioactivity community, found in any tabulation of this kind. In this study, various techniques were employed; they are described here.%
\end{abstract}%
%
%
%
%\keywords{Suggested keywords}%Use showkeys class option if keyword display desired.
%
%
%
\maketitle
\section{Introduction}%
\input{Introduction}\section{Gamma spectrometry}%
\input{gammaCounting}%
\section{ICP-MS}%
\input{icp-ms}%
\section{NAA}%
\input{naa}\section{Radon Emanation}%
This effort quantifies the amount of \rntwotwentytwo\ gas released from materials into the surrounding environment. The different methods used to acquire the results in Tab.~\ref{tab:bigtable} are described below.%
\subsection{Electrostatic counters}%
\input{rn_esc}%
\subsection{Liquid scintillation-based radon counting }%
\input{rnls}%
\section{Inter-lab comparison}%
\input{inter_lab_comp}%
\section{Data interpretation}%
\input{data_interpretation}\section{Results}%
\input{results}%
\input{bigtable2}\clearpage
\input{bigtablex}
\input{bigtable_radon}\clearpage%
\section{Conclusion}%
This paper reports the results of an extensive radioactivity study conducted in support of the planned nEXO double-beta decay experiment. The large body of data reported here can be useful for a variety of rare event searches. It will help other collaborations save time and expenses by avoiding duplication of effort. Data of the type reported here are an important enabler for a broad spectrum of science.%
\begin{acknowledgments}%
The authors gratefully acknowledge support for nEXO from the Office of Nuclear Physics within DOE’s Office of Science under grants/contracts DE-AC02-76SF00515, DE-FG02-01ER41166, DE-SC002305, DE-FG02-93ER40789, DE-SC0021388, DE-SC0012447, DE-SC0012704, DE-AC52-07NA27344, DE-SC0017970, DE-AC05-76RL01830, DE-SC0012654, DE-SC0021383, DE-SC0014517, DE-SC0024666, DE-SC0020509, DE-SC0024677 and support by the US National Science Foundation grants NSF PHY-2111213 and NSF PHY-2011948; from NSERC SAPPJ-2022-00021, CFI 39881, FRQNT 2019-NC-255821, and the CFREF Arthur B. McDonald Canadian Astroparticle Physics Research Institute in Canada; from IBS-R016-D1 in South Korea; and from CAS in China. This work was supported in part by Laboratory Directed Research and Development (LDRD) programs at Brookhaven National Laboratory (BNL), Lawrence Livermore National Laboratory (LLNL), Pacific Northwest National Laboratory (PNNL), and SLAC National Accelerator Laboratory, United States.%
\end{acknowledgments}%
\bibliography{biblio}% Produces the bibliography via BibTeX.
\end{document}

%% file: units.tex
\newcommand{\units}  {\,\ensuremath{\text{unit}}}

\newcommand{\ppq}   {\,\ensuremath{\si{ppq}}}

\DeclareSIUnit\year{yr}

\newcommand{\nm}    {\,\ensuremath{\si{nm}}}
\newcommand{\micron}{\,\ensuremath{\si{\micro m}}}

\newcommand{\mm}    {\,\ensuremath{\si{mm}}}
\newcommand{\inch}  {\ensuremath{^{\prime\prime}\mkern-1.2mu}}

\newcommand{\meter} {\,\ensuremath{\si{m}}}
\newcommand{\metre} {\,\ensuremath{\si{m}}}

\newcommand{\cmcm}  {\,\ensuremath{\si{cm^2}}}
\newcommand{\sqrm}  {\,\ensuremath{\si{m^2}}}

\newcommand{\gram}  {\,\ensuremath{\si{g}}}

\newcommand{\uBq}   {\,\ensuremath{\si{\micro\becquerel}}}

\newcommand{\keV}   {\,\ensuremath{\si{keV}}}

\newcommand{\pF}    {\,\ensuremath{\si{pF}}}
\newcommand{\nF}    {\,\ensuremath{\si{\nF}}}
\newcommand{\microF}{\,\ensuremath{\si{\micro F}}}

\newcommand{\GOhm}  {\,\ensuremath{\si{G\ohm}}}

\DeclareSIUnit\bar{bar}

%% file: isotopes.tex
\newcommand{\altwentysix}	{\ensuremath{^{26}\text{Al}}}

\newcommand{\knatural}		{\ensuremath{^\text{nat}\text{K}}}
\newcommand{\kforty}		{\ensuremath{^{40}\text{K}}}

\newcommand{\cosixty}		{\ensuremath{^{60}\text{Co}}}

\newcommand{\agonetenm}		{\ensuremath{^{110\text{m}}\text{Ag}}}

\newcommand{\csonethirtyseven}	{\ensuremath{^{137}\text{Cs}}}

\newcommand{\tltwooeight}	{\ensuremath{^{208}\text{Tl}}}

\newcommand{\pbtwoten}		{\ensuremath{^{210}\text{Pb}}}

\newcommand{\potwoten}		{\ensuremath{^{210}\text{Po}}}

\newcommand{\bitwoeleven}	{\ensuremath{^{211}\text{Bi}}}
\newcommand{\bitwotwelve}	{\ensuremath{^{212}\text{Bi}}}
\newcommand{\pbtwotwelve}	{\ensuremath{^{212}\text{Pb}}}
\newcommand{\potwotwelve}	{\ensuremath{^{212}\text{Po}}}
\newcommand{\potwofourteen}	{\ensuremath{^{214}\text{Po}}}
\newcommand{\bitwofourteen}	{\ensuremath{^{214}\text{Bi}}}
\newcommand{\pbtwofourteen}	{\ensuremath{^{214}\text{Pb}}}
\newcommand{\potwosixteen}{\ensuremath{^{216}\text{Po}}}
\newcommand{\potwoeightteen}{\ensuremath{^{218}\text{Po}}}
\newcommand{\rntwonineteen}	{\ensuremath{^{219}\text{Rn}}}
\newcommand{\rntwotwenty}	{\ensuremath{^{220}\text{Rn}}}
\newcommand{\rntwotwentytwo}	{\ensuremath{^{222}\text{Rn}}}

\newcommand{\ratwotwentysix}	{\ensuremath{^{226}\text{Ra}}}

\newcommand{\thtwotwentyeight}	{\ensuremath{^{228}\text{Th}}}
\newcommand{\actwotwentyeight}	{\ensuremath{^{228}\text{Ac}}}

\newcommand{\thtwothirtytwo}	{\ensuremath{^{232}\text{Th}}}

\newcommand{\patwothirtyfourm}	{\ensuremath{^{234\text{m}}\text{Pa}}}

\newcommand{\utwothirtyeight}	{\ensuremath{^{238}\text{U}}}
\newcommand{\nptwothirtynine}	{\ensuremath{^{239}\text{Np}}}

%% file: newcommands.tex
\newcommand{\gammacounter}{\ensuremath{\upgamma\text{ counter}}}
\newcommand{\gammacounting}{\ensuremath{\upgamma\text{ counting}}}

\newcommand{\zeronubb}{\ensuremath{0\upnu\upbeta\upbeta}}

\NewDocumentCommand\mkwidthof{O{\raggedright}mm}{
    \parbox[b]{\widthof{#2}}{#1#3}
}%
\NewDocumentCommand\mmkwidthof{O{\raggedright}mm}{
    \mkwidthof[#1]{\ensuremath{\displaystyle#2}}{\ensuremath{\displaystyle#3}}
}%

%% file: authors-nexo-radioassay2.tex
\newcommand{\UCSD}{\affiliation{Physics Department, University of California San Diego, La Jolla, CA 92093, USA}}
\newcommand{\McGill}{\affiliation{Physics Department, McGill University, Montr\'eal, QC H3A 2T8, Canada}}
\newcommand{\Stanford}{\affiliation{Physics Department, Stanford University, Stanford, CA 94305, USA}}
\newcommand{\Erlangen}{\affiliation{Erlangen Centre for Astroparticle Physics (ECAP), Friedrich-Alexander University Erlangen-N{\"u}rnberg, Erlangen 91058, Germany}}
\newcommand{\PNNL}{\affiliation{Pacific Northwest National Laboratory, Richland, WA 99352, USA}}
\newcommand{\Carleton}{\affiliation{Department of Physics, Carleton University, Ottawa, ON K1S 5B6, Canada}}
\newcommand{\UMass}{\affiliation{Amherst Center for Fundamental Interactions and Physics Department, University of Massachusetts, Amherst, MA 01003, USA}}
\newcommand{\ITEP}{\affiliation{National Research Center ``Kurchatov Institute'', Moscow, 123182, Russia}}
\newcommand{\LLNL}{\affiliation{Lawrence Livermore National Laboratory, Livermore, CA 94550, USA}}
\newcommand{\UK}{\affiliation{Department of Physics and Astronomy, University of Kentucky, Lexington, KY 40506, USA}}
\newcommand{\BNL}{\affiliation{Brookhaven National Laboratory, Upton, NY 11973, USA}}
\newcommand{\SLAC}{\affiliation{SLAC National Accelerator Laboratory, Menlo Park, CA 94025, USA}}
\newcommand{\RPI}{\affiliation{Department of Physics, Applied Physics, and Astronomy, Rensselaer Polytechnic Institute, Troy, NY 12180, USA}}
\newcommand{\Laurentian}{\affiliation{School of Natural Sciences, Laurentian University, Sudbury, ON P3E 2C6, Canada}}
\newcommand{\IHEP}{\affiliation{Institute of High Energy Physics, Chinese Academy of Sciences, Beijing, 100049, China}}
\newcommand{\IME}{\affiliation{Institute of Microelectronics, Chinese Academy of Sciences, Beijing, 100029, China}}
\newcommand{\Sherbrooke}{\affiliation{Universit\'e de Sherbrooke, Sherbrooke, QC J1K 2R1, Canada}}
\newcommand{\Alabama}{\affiliation{Department of Physics and Astronomy, University of Alabama, Tuscaloosa, AL 35487, USA}}
\newcommand{\UNCW}{\affiliation{Department of Physics and Physical Oceanography, University of North Carolina Wilmington, Wilmington, NC 28403, USA}}
\newcommand{\UBC}{\affiliation{Department of Physics and Astronomy, University of British Columbia, Vancouver, BC V6T 1Z1, Canada}}
\newcommand{\TRIUMF}{\affiliation{TRIUMF, Vancouver, BC V6T 2A3, Canada}}
\newcommand{\Drexel}{\affiliation{Department of Physics, Drexel University, Philadelphia, PA 19104, USA}}
\newcommand{\ORNL}{\affiliation{Oak Ridge National Laboratory, Oak Ridge, TN 37831, USA}}
\newcommand{\CSU}{\affiliation{Physics Department, Colorado State University, Fort Collins, CO 80523, USA}}
\newcommand{\Yale}{\affiliation{Wright Laboratory, Department of Physics, Yale University, New Haven, CT 06511, USA}}
\newcommand{\USD}{\affiliation{Department of Physics, University of South Dakota, Vermillion, SD 57069, USA}}
\newcommand{\Mines}{\affiliation{Department of Physics, Colorado School of Mines, Golden, CO 80401, USA}}
\newcommand{\CUP}{\affiliation{IBS Center for Underground Physics, Daejeon, 34126, South Korea}}
\newcommand{\UWC}{\affiliation{Department of Physics and Astronomy, University of the Western Cape, P/B X17 Bellville 7535, South Africa}}
\newcommand{\SUBATECH}{\affiliation{SUBATECH, Nantes Universit\'e, IMT Atlantique, CNRS/IN2P3, Nantes 44307, France}}
\newcommand{\Caltech}{\affiliation{California Institute of Technology, Pasadena, CA 91125, USA}}
\newcommand{\Bern}{\affiliation{LHEP, Albert Einstein Center, University of Bern, 3012 Bern, Switzerland}}
\newcommand{\Skyline}{\affiliation{Skyline College, San Bruno, CA 94066, USA}}
\newcommand{\SNOLAB}{\affiliation{SNOLAB, Lively, ON P3Y 1N2, Canada}}
\newcommand{\Muenster}{\affiliation{Institut f{\"u}r Kernphysik, Westf{\"a}lische Wilhelms-Universit{\"a}t M{\"u}nster, M{\"u}nster 48149, Germany}}
\newcommand{\FRIB}{\affiliation{Facility for Rare Isotope Beams, Michigan State University, East Lansing, MI 48824, USA}}
\newcommand{\Queens}{\affiliation{Department of Physics, Queen's University, Kingston, ON K7L 3N6, Canada}}
\newcommand{\Windsor}{\affiliation{Department of Physics, University of Windsor, Windsor, ON N9B 3P4, Canada}}
\newcommand{\TUM}{\affiliation{Physikdepartment and Excellence Cluster Universe, Technische Universit{\"a}t M{\"u}nchen, Garching 80805, Germany}}
\newcommand{\Montclaire}{\affiliation{Department of Physics and Astronomy, Montclair State University, Montclair, NJ 07043, USA}}
\newcommand{\McMaster}{\affiliation{Department of Physics and Astronomy, McMaster University, Hamilton, ON L8S 4M1, Canada}}
\newcommand{\SJTU}{\affiliation{School of Physics and Astronomy, Shanghai Jiao Tong University, Shanghai 200240, China}}
\newcommand{\Hawaii}{\affiliation{Department of Physics and Astronomy, University of Hawaii at Manoa, Honolulu, HI 96822, USA}}
\newcommand{\KEK}{\affiliation{KEK, High Energy Accelerator Research Organization 1-1 Oho, Tsukuba, Ibaraki 305-0801, Japan}}

%Collaboration authors
\author{R.~MacLellan\textsuperscript{\orcidlink{0000-0003-2479-5277}}}\UK
\author{P.~Acharya}\Alabama
\author{B.~Aharmim}\Laurentian  % not nEXO author
\author{S.~Alcantar Anguiano}\PNNL %not nEXO author
\author{A.~Anker}\SLAC
\author{I.~J.~Arnquist\textsuperscript{\orcidlink{0000-0002-5643-8330}}}\PNNL
\author{D.~Auty}\altaffiliation{Now at: University of Alberta, Department of Physics, Canada}\Alabama % not nEXO author
\author{T.~Bhatta}\altaffiliation{Now at: Cleveland Community College, USA}\UK
\author{D.~Chernyak\textsuperscript{\orcidlink{0000-0001-6162-3453}}}\altaffiliation{Now at: Research Center for Neutrino Science, Tohoku University, Sendai 980-8578, Japan}\Alabama
\author{J.~S.~Choe}\CUP  % not nEXO author
% ...

%% file: Introduction.tex
Access to materials containing the smallest possible amount of radioactivity is important for a number of scientific and technical applications. 
Random energy deposits arising from minute amounts of primordial, cosmogenic, and man-made radioactivity in materials can limit the sensitivity of radiation detectors used in scientific, health, and national security applications.
Unwanted radioactivity in materials can lead to quantum decoherence in superconducting quantum circuits~\cite{Loer_2024}.
The ever-present naturally-occurring radioisotopes \kforty, \thtwothirtytwo, and \utwothirtyeight\ and associated radioactive decay progeny are 
of particular concern for a broad range of applications. Below, we discuss a range of applications that require low-radioactivity materials.
Searches for phenomena forbidden within the framework of the Standard Model of Particle Physics (SM) are a tool for the discovery of new physics, often without requiring large particle accelerators. Such experiments often play out at keV or MeV energy scales and involve searching for exceedingly rare phenomena. 
Because of this combination of circumstances, the avoidance of background events arising from radioactivity in and on the materials comprising the apparatus is an important and often time-consuming aspect of experiment planning, design, and construction. Radioactivity-induced background events often limit the achievable sensitivity by masking the signature of the phenomena of interest.

Searches for neutrinoless double-beta decay (\zeronubb) are a means to explore physics beyond the SM, and they require radiopure materials. The relevant energy for \zeronubb\ is on the order of a few MeV~\cite{RevModPhys.95.025002}.
Detection of this never-observed ultra-rare nuclear decay mode would demonstrate that neutrinos and anti-neutrinos are the same species, requiring physics beyond the SM. To be scientifically relevant, future searches aim for half-life sensitivities of $10^{27}$ to $10^{28}$ yr. At this level of stability, tonnes of decaying material will undergo only a few decays per year. Given the rarity of these decays, experiments are utilizing various techniques for active background suppression. However, as a starting point, all require the use of structural and technical components containing minimal amounts of radioactivity
in order to not overwhelm them. To meet their extreme radiopurity requirements, proposed \zeronubb\ experiments such as nEXO~\cite{Adhikari_2022}, LEGEND-1000~\cite{legendcollaboration2021legend1000preconceptualdesignreport}, and CUPID~\cite{CUPID:2019imh}, conduct extensive materials screening campaigns to identify materials and components.

The work reported here was performed in support of the technical development of the planned nEXO experiment.
This paper reports on the assay data for materials of interest to nEXO~\cite{Adhikari_2022}. However, the reported data are of broad interest. 
Given its connection to nEXO, the paper briefly outlines the tools developed by this project to interpret the large body of radioactivity data and place it within the framework of a background rate compilation. This approach allows for real-time materials decision-making.

Material radioactivity compilations, similar to this one, have been published by various solar neutrino, double-beta decay and dark matter searches---BOREXINO~\cite{ARPESELLA20021}, EDELWEISS~\cite{ARMENGAUD20131}, EXO-200~\cite{LEONARD2008490,LEONARD2017169}, GERDA~\cite{BUDJAS2009755}, LUX~\cite{AKERIB201533}, LZ~\cite{Akerib2020}, Majorana Demonstrator~\cite{ABGRALL201622}, NEXT-100~\cite{Alvarez_2013}, SNO~\cite{JAGAM1993389}, XENON100~\cite{APRILE201143}, XENON1T~\cite{Aprile2017}, and XMASS~\cite{Abe_2020}---to document the body of knowledge in selecting suitable low-radioactivity materials,  which has been accumulated by the low background community. These compendia of material assay results provide future experiments with a general assessment of what is currently achievable with various low-radioactivity materials, an essential input for current and future low-background applications.

To provide useful constraints and driven by the technical complexity of modern experimental designs, numerous radioactivity measurements must be performed at the level of \si{\micro Bq}/kg or even nBq/kg across a variety of materials. Such material assays are technically challenging.  

Two approaches are typically chosen to meet this analytical challenge: decay counting and atom counting.
These approaches differ in what they determine and in the analytical sensitivity they reach.
For experiments where backgrounds created by energetic $\upbeta$ and $\upgamma$ radiation are a concern, e.g., double-beta decay searches, \thtwothirtytwo\ and \utwothirtyeight\ progeny are important background sources. These are primordial radioactive isotopes that have been present since before the formation of our solar system. These nuclides have half-lives of \SI{14.0e9}{\year} and \SI{4.468e9}{\year}, respectively, and are typically found in all materials at trace levels. To meet this analytical challenge, the effort described here utilized both approaches.

\begin{enumerate}
   \item Decay counting is based on the determination of the \kforty, \actwotwentyeight, \pbtwotwelve, \bitwotwelve, \tltwooeight, \patwothirtyfourm, \ratwotwentysix, \pbtwofourteen, \bitwofourteen\ decay rates. The conversion of activities into background event rates is often independent of assumptions about decay-chain equilibrium. While non-destructive, this approach typically offers limited sensitivity (tens of ppt for Th and U, or roughly the \SI{100}{\uBq/\kilogram} range), requires large samples (kg-scale) and is time-consuming (2--3 weeks per unique sample).\\
   The following techniques were used in this work: low background $\upalpha$ and $\upgamma$ spectrometry and radon counting.
   \item The atom counting approach for \knatural, \thtwothirtytwo, and \utwothirtyeight\ only requires limited amounts of samples (g-scale and smaller), offers fast turn-around (days to weeks per unique sample) and has excellent sensitivity (sub-ppt, equating to \si{\micro\becquerel/\kilogram} or sub-\si{\micro\becquerel/\kilogram}), but is destructive. 
   This approach is generally sensitive to much lower activities than decay counting, but is only sensitive to the top of the \thtwothirtytwo\ and \utwothirtyeight\ decay chains.\\
The following techniques were used in this work: inductively coupled plasma mass spectrometry (ICP-MS), glow discharge mass spectrometry (GD-MS), and neutron activation analysis (NAA).
\end{enumerate}
This work exploits both approaches to meet the analytical challenge. 
%
%Atom counting is used whenever its superior sensitivity is required.
%
If decay-chain equilibrium is assumed, the high sensitivity of atom counting can be used to improve constraints on the lower-chain activities.  However, assumptions about decay-chain equilibrium breakage need to be evaluated on a case-by-case basis.
%
%The question of whether decay-chain equilibrium is broken needs to be evaluated on a case-by-case basis. 
This evaluation must account for the chemical and physical properties (electrochemical potential, melting point, boiling point, possible oxidation states) of the material under study, as well as the associated naturally occurring radioisotopes with half-lives on or longer than the timescales relevant to the experiment and the processes. The processes used to purify the material also need to be considered. Equilibrium breakage is of particular importance in the \utwothirtyeight\ decay series. \ratwotwentysix\ and \pbtwoten\ can decay at higher or lower rates than \utwothirtyeight. The chemical or physical process responsible for the imbalance determines the type and extent of fractionation among decay-chain isotopes. As an example, the EXO-200 experiment applied such an equilibrium assumption to all ICP-MS and NAA measurements; the pre-data-taking background rate predictions were later found to overlap with the data-derived rates~\cite{PhysRevC.92.015503}; in this instance, the success of that experiment justified the approach and the assumptions made about decay-chain equilibrium post hoc.

Thus, the choice of assay method must strike a balance between sensitivity and the risk of background misestimation due to decay-chain equilibrium breakage. Currently, no analysis approach that addresses both issues simultaneously exists.

Various assay installations are used in support of the study described here. They are operated by nEXO collaborating or associated institutions, some are above ground, some underground.

The sections below provide a summary of the tools, followed by an extensive compilation of results.

%% file: gammaCounting.tex
\newcommand{\gei}  {Ge-I}%
\newcommand{\geii} {Ge-II}%
\newcommand{\geiii}{Ge-III}%
\newcommand{\geiv} {Ge-IV}%
\begin{table*}%
    \caption{\label{tab:ge}Overview of the properties\footnote{The stated efficiency of germanium detectors relative to the absolute full-energy peak efficiency of a \qty{7.6}{cm} diameter by \qty{7.6}{cm} height NaI(Tl) crystal to the \qty{1.33}{\mega\eV} peak of a \cosixty\ source positioned \qty{25}{cm} from the detector, or \qty{1.2e-3} absolute efficiency.} of HPGe detectors used to assay samples as presented in this work. All detectors are p-type. All detectors are coaxial, except for CWELL, which is a well detector. Data were provided by the detector operators. A description of GeMPI 2 has been published in~\cite{HEUSSER2006495,ACKERMANN2023110652}. Relative efficiencies are based on original manufacturer-provided values. The background rates are quoted for various peak energies and are per unit mass of the Ge crystal.} %
    \begin{ruledtabular}%
        \begin{tabular}{l c S S S S l}%
            Name            &%
            Efficiency      &%
            \multicolumn{4}{c}{Background rate in counts/(d$\cdot$kg)} &%
            Location\\
            \cline{3-6}
                            &
            (rel $\%$)      &
             {609\keV}      &
            {1332\keV}      &
            {1461\keV}      &
            {2615\keV}\\
            \hline
            \geii           & 
            \phantom{0}60   & 
            <7.6            & 
            10              &
            2.3             &
            0.75            &
            UA\\
            \geiii          & 
            105             &  
            9.6             & 
            1.7             &
            2.5             &
            <0.53           &
            UA\\
            \geiv           &
            111             &
            0.33            &  
            0.48            &
            0.19            &
            0.25            &
            SURF\\
            %
            %Mordred         & 
            %\phantom{0}60   &  
            %3.9             & 
            %1.6             &
            %7.4             &
            %2.1             &
            %SURF\\
            %
            PGT             & 
            \phantom{0}55   &  
            0.58            & 
            0.22            &
            1.4             &
            0.50            &
            SNOLAB\\
            CWELL           & 
            \phantom{0}80   &  
            0.041           & 
            <0.009          &
            <0.008          &
            <0.007          &
            SNOLAB\\
            Lively          & 
            107             &  
            1.03            &
            0.118           &
            0.57            &
            0.94            & 
            SNOLAB\\
            Gopher          & 
            120             &  
            3.0             & 
            0.085           &
            0.30            &
            0.050           &
            SNOLAB\\
            VdA             & 
            120             &  
            1.4             & 
            0.059           &
            1.3             &
            0.51            &
            SNOLAB\\
            GeMPI 2         & 
            100             & 
            <0.12           & 
            <0.11           &
            <0.13           &
            <0.10           &
            LNGS %
        \end{tabular}%
    \end{ruledtabular}%
\end{table*}%

High-resolution $\upgamma$-ray spectroscopy with high-purity germanium detectors (HPGe), or \gammacounting, provides a direct measurement of the Th and U progeny responsible for the emission of energetic, MeV-scale $\upgamma$ radiation. In particular, \gammacounting\ is directly sensitive to decays of \tltwooeight\ and \bitwofourteen, isotopes in the natural radioactive decay chains of \thtwothirtytwo\ and \utwothirtyeight, respectively. In cases where the breakage of secular equilibrium exists, \gammacounting\ is the only way to determine the decay rates of the progeny, but with limited sensitivity.

High-sensitivity, low-background \gammacounter s are located deep underground, where they are shielded from cosmic rays. Given the limited availability of underground \gammacounter s, less-sensitive above-ground gamma counters are used to pre-screen high-purity materials or to analyze materials with less stringent radiopurity requirements.

Measurements presented here were obtained using two \gammacounter s located in a surface laboratory at the University of Alabama (UA). The specifications of these counters are listed in Table \ref{tab:ge} under \geii\ and \geiii. Both are equipped with anti-coincident plastic scintillator cosmic-ray muon detectors with almost 4$\pi$ coverage~\footnote{Some older measurements were taken with a less efficient muon detection system, prior to it being upgraded.}. The sensitivity of these \gammacounter s is limited by background from neutron-induced delayed-coincidence photons. In addition, some of the data were obtained by seven underground gamma counters. \geiv, operated by nEXO collaborators, is located on the 1500\metre\ level of the Sanford Underground Research Facility (SURF). PGT, CWELL, Lively, Gopher, and Vue-des-Alpes (VdA) are operated by the SNOLAB Low Background Counting Facility\cite{Lawson:2023Ep} located 2100\metre\ underground. The VdA counter, 80\% of which was dedicated to the nEXO Project, was initially operated 620\metre\ underground in the Vue-des-Alpes underground laboratory in Switzerland. One of the measurements presented here was performed with the GeMPI 2 counter at the INFN Gran Sasso National Laboratory (LNGS), 1400\metre\ underground. The number of assays performed with the \geiv\ and VdA counters was limited by extended periods during which these detectors were not operational. %

%% file: icp-ms.tex
ICP-MS can identify and, with varying sensitivities, quantify almost all isotopes in the periodic table of nuclides, including K, Th, and U.
The technique is based on the analysis of the distribution of ion momenta of a sample.
ICP-MS is a destructive technique and typically requires the sample to be dissolved in acid before being introduced, as a fine-mist aerosol via a nebulizer, into the mass spectrometer using a stream of argon gas. In a plasma, ions are formed and extracted into a high-vacuum region, where, using ion lenses, they are focused into a beam and directed to the mass analyzer. The analyte components are then analyzed based on their mass-to-charge ratio. Calibration standards are used for quantitation either by using an external calibration curve, standard addition, or isotope dilution methods. The latter is the most accurate and precise method, applied to most of the ICP-MS assays reported here.

Because ICP-MS can detect $^{232}$Th, $^{238}$U, and natural K with high sensitivity when coupled with ultra-clean sample preparation procedures, it is widely employed in the realm of low-background experiments for detector material validation, e.g., \cite{LAFERRIERE201593, ARNQUIST2020163761, ARNQUIST2020163573, Arnquist2023,D0JA00220H,diVacri2022,Arnquist2018}. 
Quantities of material on the order of a gram or a fraction of a gram are required to perform assays at sensitivities relevant to the ultra-low-background community. 
The advantage of analyzing very small amounts, however, raises the question of how representative the analyzed mass is of a large mass of the material, and how homogeneity can be verified. In this work, whenever possible, ICP-MS analysis of multiple sub-samples was performed for materials intended to be used in large masses, where homogeneity could not be guaranteed.
An industry-inspired approach to part sampling, as described in~\cite{ANSI_sampling_2018}, is planned for implementation in nEXO. 
Analysis times are on the order of days, including the sample preparation, which is the most time-consuming aspect of the measurement process. The instrumental analysis requires only minutes, and multiple elements can be monitored in a single run. For the results reported here, sensitivities in the parts-per-trillion (ppt) and sub-ppt ranges have been routinely achieved for the assay of Th and U in materials. The best sensitivities for \thtwothirtytwo\ and \utwothirtyeight\ determinations were obtained for a sample of heat-transfer fluid at 1.3 parts per quadrillion (ppq), or \qty{5.3}{\nano\becquerel/\kilogram}, and 4.3\ppq, or \qty{53}{\nano\becquerel/\kilogram}, respectively. Three collaborating facilities contributed ICP-MS capability to this study: Pacific Northwest National Laboratory (PNNL) in the USA, the IBS Center for Underground Physics (CUP) in Korea, and the Institute of High Energy Physics (IHEP) in China.

In addition to analyses of bulk radiopurity, ICP-MS can be leveraged to optimize and validate material manufacturing processes and cleaning procedures owing to its high sensitivity, relatively short analysis time, and rapid sample throughput via batching. In the work reported in \cite{ARNQUIST2020163573,Arnquist2023}, over 300 unique assays were performed to understand the contamination of complex part processing, namely flexible cable manufacturing. These investigations allow for the ability to isolate the contaminating steps, to develop mitigating steps toward clean processing, and/or to develop surface cleaning techniques to remove contamination from surfaces. %

%% file: naa.tex
Neutron activation analysis (NAA) is a well-proven, ultra-sensitive technique for trace-element detection and quantification. It involves irradiating a sample in a nuclear reactor, thereby transmutating stable or long-lived radioactive nuclides to those with comparatively short half-lives, such as \utwothirtyeight\ to \nptwothirtynine. It can be used to detect chemical impurities with large neutron-capture cross sections and activation products with sufficiently long lifetimes. Knowledge of the neutron flux, activation duration, and capture cross sections allows for the conversion of measured activation-product activities into concentrations of the parent element. Aside from careful surface cleaning, no chemical sample preparation is required. Typical sample mass is on the order of a few grams. As with ICP-MS, it measures the long-lived heads of the natural decay series. The NAA effort is conducted by the UA group, using the MIT research reactor MITR as the neutron source. For \thtwothirtytwo\ and \utwothirtyeight, ppt-sensitivities have been routinely achieved, with sub-ppt results reached in exceptional cases.

The EXO-200 double beta decay experiment made extensive use of NAA in their materials assay campaign \cite{LEONARD2008490,LEONARD2017169}. Apart from the improvements described below, the work presented here largely uses the same technique as EXO-200. For an overview of the method, see the two EXO-200 Refs.~\cite{LEONARD2008490,LEONARD2017169}.

Compared to EXO-200, a new analysis code was developed for NAA data. Analysis is done with a system of ROOT macros, simultaneously fitting peaks for multiple activities, similar to the EXO-200 code \cite{10.3389/fphy.2024.1362209}, which is still under active development. The ROOT macros have high automation and can fit many peaks in many spectra rapidly. Based on the significance of the candidate peak, it automatically finds peaks in $\upgamma$ spectra. The automation features enable the rapid analysis of multiple isotopes.  

To increase sensitivity for \utwothirtyeight, $\upgamma\upgamma$-coincidence counting can be used. This approach makes use of correlated $\upgamma$ emissions observed in the level scheme of \nptwothirtynine.
A computational study showed, for example, that this method can enhance U-sensitivity by a factor of 8 for the activation analysis of sapphire~\cite{Tsang_2021}, where large side-activities limit traditional NAA sensitivity. A measurement with two HPGe detectors, performed on an activated sapphire sample (\# \ref{R-048.6.2} in Tab.~\ref{tab:bigtable}), demonstrated the validity of the method when compared with a previous assay (\# \ref{R-048.6.1} in Tab.~\ref{tab:bigtable}) that did not use coincidence counting~\cite{zgzm-w5f4}.

%% file: rn_esc.tex
Electrostatic counters (ESCs) collect ionized radon progeny that decay in the counter vessel and are electrostatically guided onto an alpha detector, where subsequent decays can be measured. Instruments operated by Laurentian University were initially built to assay radium content in water for SNO. These are described in \cite{WANG1999601,ANDERSEN2003399,10.1063/1.2060472}. Similar instruments at SLAC National Accelerator Laboratory (SLAC) are newly constructed by nEXO; these instruments are described by \citet{ANKER2026170876}. Both suites of instruments operate under the same principle.  

The material to be assayed is placed in a gas recirculation loop with the ESC and a pump. The loop is pumped out, and a high-purity carrier gas (Ar/$\mathrm{N_2}$/He/Xe) is filled into the system at pressures between \SI{20}{\milli\bar} and \SI{1}{\bar}, depending on the compatibility of the sample and goal of the measurement. The pump recirculates the gas in the system, carrying the radon atoms from the sample to the electrostatic vessel. A large-area silicon photodiode in the ESC is biased with negative high-voltage to attract ionized progeny where they plate out, and subsequent $\upalpha$ decays in the series may be detected. Decays that deposit energy on the photodiode create a proportional amount of charge signal that is amplified and recorded, along with the time of the event. Since the sample is connected to the ESC during the measurement, it is possible to observe all three natural decay series containing radon---\rntwotwentytwo\ with $T_{1/2}=\SI{3.8222}{\day}$, \rntwotwenty\ with $T_{1/2}=\SI{55.6}{\second}$, and \rntwonineteen\ with $T_{1/2}=\SI{3.96}{\second}$---using these instruments. Lower carrier gas pressures reduce the transport time for a given mass flow, thereby delivering the short-lived radon isotopes to the counting chamber before they decay.

The data analysis uses the $\upalpha$ energies and times---from the decay of \potwoeightteen, \potwosixteen, \potwofourteen, \potwotwelve, \bitwotwelve, and \bitwoeleven---to fit to the Bateman equations for the various radon decay series to determine the rate of radon emanation. The detectors are calibrated using a \rntwotwentytwo\ source to determine the overall detection efficiency, which scales the counts detected to the actual rate of emanation. Detection sensitivities of \SI{20}{\micro\becquerel} (supported steady-state activity) have been observed. The analysis and calibration are described in more detail by \citet{ANKER2026170876}. %

%% file: rnls.tex
The UA radon counting facility measures the radon emanation rate of material samples by counting delayed coincidences of $^{214}$Bi $\upbeta$ decay and $^{214}$Po $\upalpha$ decay (Bi-Po events) from $^{222}$Rn dissolved in organic liquid scintillator (LS). 

A sample is first placed in an electropolished stainless-steel emanation chamber to allow radon to emanate into a boil-off nitrogen atmosphere for two weeks. The nitrogen, containing any emanated radon, is then transferred to a volume of LS using boil-off nitrogen as the carrier gas. A sparger promotes the transfer of the radon from the gas into the LS. The LS used is a mixture of dodecane (C$_{12}$H$_{26}$, 80\%\,vol), pseudocumene (1,2,4-trimethylbenzene, 20\%\,vol), and \SI{1.5}{\gram/\liter} of PPO (2,5-diphenyloxazole).
After bubbling the carrier gas through the LS, the LS is transferred to an acrylic counting cell, which is viewed by a Hamamatsu R1307 photomultiplier tube. Counting occurs in a lead-shielded dark box within a stainless steel enclosure to reduce the effects of external $\upgamma$ radiation and light.

Coincident $^{214}$Bi and $^{214}$Po decays are identified by the amount of light detected, which indicates the energy of the decays, and the time between two consecutive light detections. The time difference is expected to follow an exponential distribution with a decay constant corresponding to the mean lifetime of \potwofourteen\ if the signals are due to Bi and Po decays. 
Data taking is divided into runs, each lasting approximately three days. An exponential fit to the coincidence times determines the radon concentration in the cell for each run. An exponential fit to all runs, where the individual rates are expected to diminish with the mean lifetime of $\rm ^{222}Rn$, is performed to determine the radon-induced rate at the time of loading. After accounting for detection and transfer efficiencies, the radon emanation rate of the sample is calculated. A detailed description of this approach is provided by \citet{Sazzad_2025}. A 90\% CL detection limit of \SI{1.6}{\milli\becquerel} and minimal detectable activity of \SI{3.0}{\milli\becquerel} (both in terms of the supported steady state activity) have been reported by \citet{Sazzad_2025}. %

%% file: inter_lab_comp.tex
To ensure the reliability of the dataset, the correctness and consistency of the measurement results need to be verified. This is particularly important for a measurement program that relies on data collected across multiple laboratories. The quality control program described in this paper relied on comparative sample measurements. %

\input{pgf}%
\definecolor{tabred}{HTML}{d62728}%
\definecolor{tabgreen}{HTML}{2ca02c}%
\definecolor{tabblue}{HTML}{1f77b4}%
\definecolor{taborange}{HTML}{ff7f0e}%
\begin{figure*}[t]%
  \centering
  \subfloat[\label{fig:nondest}\raggedright Non-destructive testing of standards containing primarily \kforty\ ({\color{tabblue}\tabblackDiamond[0.6]}), \thtwotwentyeight\ ({\color{tabred}\tabblackcircle[0.6]}), and \utwothirtyeight\ ({\color{tabgreen}\tabblacksquare[0.6]}). Pairs of points represent two samples of the same standard. Horizontal bars indicate the 1$\sigma$ calibrated activity of the standards. \kforty\ is scaled by a factor of 0.5.]%
  {\includegraphics[width=0.47\textwidth]{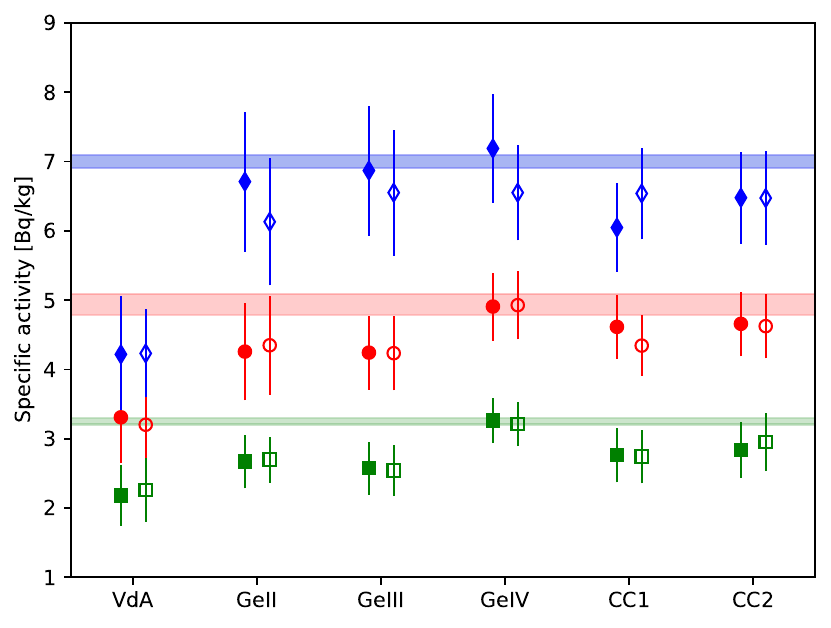}}%
  \hfill%
  \subfloat[\label{fig:dest}\raggedright Destructive testing of %
  \cref{R-093.1.1,R-093.1.2,R-093.1.3} ({\color{tabred}\tabblackcircle[0.6]}), %
  \cref{R-096.1.1,R-096.1.2} ({\color{tabblue}\tabblacksquare[0.6]}), %
  \cref{R-076.1.2,R-076.2.1,R-076.2.2} ({\color{tabgreen}\tabblackDiamond[0.6]}), and %
  \cref{R-168.1.1} ({\color{taborange}\tabblacktriangleright[0.6]}). %
  Open markers are \thtwothirtytwo; solid markers are \utwothirtyeight. %
  Horizontal lines indicate averages where appropriate. %
  Only one of the samples could be analyzed via NAA.]%
  {\includegraphics[width=0.49\textwidth]{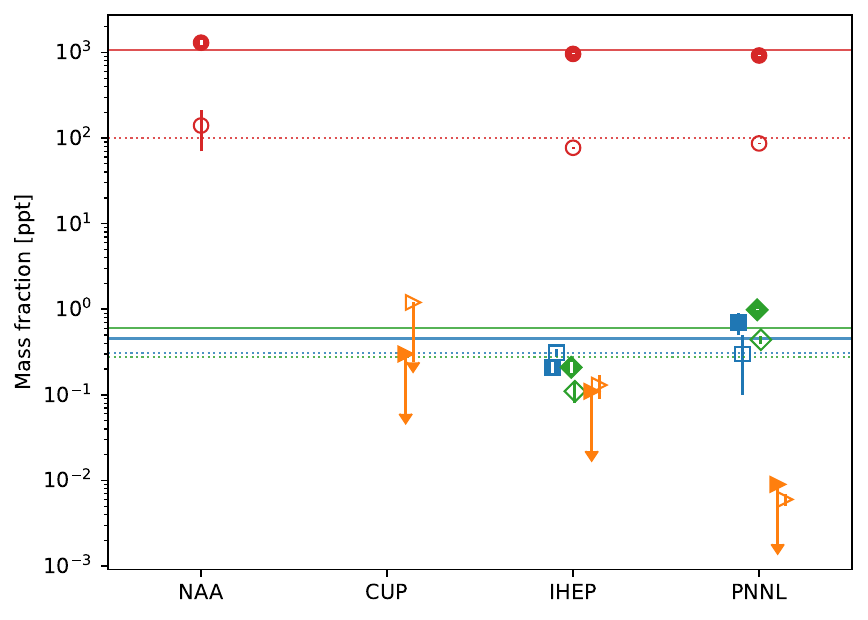}}%
  \caption{\label{fig:comparison}Assay results from six HPGe detectors---VdA underground at SNOLAB, GeII and GeIII above ground at UA, GeIV underground at SURF, and CC1 and CC2 underground at Y2L---for three pairs of calibrated standards; and the results from three samples using destructive assay techniques---NAA at UA and ICP-MS at CUP, IHEP, and PNNL---of four samples.}%
\end{figure*}%

The compatibility and correctness of the Ge detector results were tested by counting six samples, prepared at Laurentian University from activity-certified IAEA material \cite{iaea1987}. The activity of the samples was not known to the analyzers. All samples were counted, in turn, using six low-background Ge detectors located at UA, and underground at SNOLAB, SURF, and Yangyang Underground Laboratory (Y2L), operated by CUP. Excluding the VdA detector at SNOLAB, which experienced reliability issues at the time of the study, the average ratio of measured to design activities was \qty[multi-part-units=single]{0.89+-0.07} (standard deviation). The average was taken over all detectors and all activities. Fig.~\ref{fig:nondest} shows the individual specific activities for all detectors and radionuclides. The horizontal lines indicate the design-specific activity of the samples; the shaded bands their uncertainty. The level of agreement is judged to be acceptable for this purpose. %It is interesting to note that the \geiv\ results, showing the best agreement, were the only ones based on a simulation accounting for coincidence summing of the various $\upgamma$ transitions.

The comparative study of ICP-MS and NAA capabilities used low-activity materials: electroformed Cu, a P-doped Si wafer, a processed Si wafer, and Kapton. Due to the low activity requirement, it did not benefit from activity-certified standard materials which are not available at these low-activity levels. Therefore, it only tested assay compatibility between the participating labs, within the heterogeneity of the samples tested. Measurements performed at UA using NAA and at Y2L, IHEP, and PNNL using ICP-MS are compared.

\begin{table}%
\caption{\label{tab:destructive}Mean and standard deviation of measurements of samples for destructive testing comparison. Measured limits omitted from the mean.}
  \begin{ruledtabular}%
    \begin{tabular}{lS[table-format=3.3+-2.3]S[table-format=4.2+-3.2]}%
      Sample (\# in \cref{tab:bigtable}) & {\thtwothirtytwo/ppt} & {\utwothirtyeight/ppt}\\
      \hline
      Kapton (\ref{R-093.1.1} to \ref{R-093.1.3})  & 101+-34 & 1060+-210\\
      P-doped Si (\ref{R-096.1.1} and \ref{R-096.1.2})          & 0.305+-0.007	& 0.46+-0.35\\
      Processed Si (\ref{R-076.1.2} to \ref{R-076.2.2}) & 0.28+-0.23 &	0.5+-0.6\\
      Copper (\ref{R-168.1.1})                    & 0.07+-0.09 \\
    \end{tabular}%
  \end{ruledtabular}%
\end{table}%
Fig.~\ref{fig:nondest} shows the distribution of analysis results. Due to interfering side activities, NAA could only be performed on the Kapton sample. As expected, due to the smallness of the concentrations tested, the variability is greater than observed for Ge counting. %
The average, with standard deviation, measured activity of the four test samples are listed in \cref{tab:destructive}. The IHEP measurement of \thtwothirtytwo\ in the copper sample is likely anomalous. The PNNL measurement employs an efficiency tracer and consistently measures ppq levels of uranium and thorium in electroformed copper, implying that the IHEP sample suffered from some external contamination rather than a fundamental limit to their sensitivity. We highlight this discrepancy to emphasize the need for repeated, competitive testing of important samples with such low levels of residual contamination.

%% file: data_interpretation.tex
The radioactivity data set reported here has intrinsic value. It is useful for multiple projects and differing scientific applications. However, it is instructive to put the data in some context. Such context is provided by the procedures put into place by the nEXO Collaboration, designed to interpret the radioactivity data and answer the question ``what is good enough''? An online database, with associated applications, can be used to convert specific activities into background rates in real time~\cite{TSANG2023168477}. Its associated applications enable material acceptance decisions, formalize material analysis requests, communicate assay completion, and enable background rate accounting between major simulation updates. An application enables simplified acceptance decisions using a web browser and the database content.

Coupled with GEANT4-based detector simulations, also stored in the database, the radioactivity data can be converted into component-specific background event rates for various event classes, fiducial detector masses, and energy windows. The results are presented as a large spreadsheet, which serves as the interface to the engineering component of the project. Component-specific probability density functions (PDFs) are stored in the same database and then used to convert decay rates into event rates. All non-linear aspects of the background estimation are contained in the PDFs. 
The database and its associated applications, in contrast, aggregate the product of component mass, specific activities, and hit efficiencies as a linear combination of all component background rates.

During development, the experiment design is evolving. For nEXO, significant upgrades to the detailed detector and background models were made infrequently, primarily for major publications or reviews. The database bookkeeping function allows for continuous development of the background model between simulation updates. This is done by assigning hit-efficiency proxies to new detector regions and incorporating them into an approximate background model. It has been found that the background rates of major model updates are approximated reasonably well by the running model. This approach enables real-time decision-making on new detector components during the design phase, and not only when major model updates occur. It leaves documentation in the form of the computer-generated background rate spreadsheet summary.

The database and associated suite of applications allow nEXO to identify the leading background contributors in real time. Within nEXO, this capability has been used to assign assay priorities and manage the various analysis laboratories. The database is online, and its various capabilities are available to all collaborators. The question of ``what is the current background rate'' can be answered at any time and is entirely data-driven. %

%% file: results.tex
\newcommand{\ff}{\ensuremath{^\sim}}
Hundreds of materials have been measured for the R\&D of nEXO using the techniques described above.
The results of these measurements have been stored in the nEXO Materials Database \cite{TSANG2023168477}.
The assay results for early chain \utwothirtyeight, late chain \utwothirtyeight\ (\ratwotwentysix), \thtwothirtytwo, and \kforty\ are shown in Tab.~\ref{tab:bigtable}. Results for other isotopes, corresponding to entries in Tab.~\ref{tab:bigtable} that are flagged with a $^\dagger$ following their number, are shown in Tab.~\ref{tab:bigtablex}. Radon counting results, together with some from EXO-200 that had not yet been published, are shown in Tab.~\ref{tab:bigtableradon}. Results are shown following the precision and rounding convention of Ref.~\cite{PhysRevD.110.030001}. All uncertainties are $1\sigma$. Limits are calculated following the prescription of \citet{PhysRevD.110.030001} at the 90\%\,C.L., except for results that are less than $-3\sigma$, for which the limit shown is $1.645\sigma$. These results, all in Tab.~\ref{tab:bigtablex}, are indicated with a \ff. For many results presented as limits, the central value and associated uncertainty are available upon request. 
To save space, abbreviations have been used in Tab.~\ref{tab:bigtable} and Tab.~\ref{tab:bigtableradon}.
Descriptions of the abbreviations used are provided in Tab.~\ref{tab:abbr}.%
\input{abrev}

%% file: abrev.tex
\begingroup%
    \squeezetable%
    \begin{table}[h]%
        \caption{\label{tab:abbr}Table of abbreviations used in Tab.~\ref{tab:bigtable} and Tab.~\ref{tab:bigtableradon}}%
        \begin{ruledtabular}%
            \SetTblrInner{rowsep=0pt}
            \begin{tblr}{width=\columnwidth, colspec={l X}, colsep=2pt}%
                Abbreviation & Description\\
                \hline%
		A.G.~Ge\label{acr:agge}  & Above Ground Germanium Counting                           \\
		APT\label{acr:apt}       & Advanced Polymer Technologies (Now MCAM)                  \\
		AWG\label{acr:awg}       & American Wire Gauge                                       \\
		ASIC\label{acr:asic}     & Application-specific Integrated Circuit                   \\
		BNL\label{acr:bnl}       & Brookhaven National Laboratory                            \\
        BP\label{acr:bp}         & Boedeker Plastics                                         \\
        CAS\label{acr:casrn}     & Chemical Abstracts Service Registration Number            \\
		CMOS\label{acr:cmos}     & Complementary Metal-Oxide-Semiconductor                   \\
		CUP\label{acr:cup}       & IBS Center for Underground Physics                 \\
		CVD\label{acr:cvd}       & Chemical Vapor Deposition                                 \\
                %CVMR: trademarked name
        DP\label{acr:dp}         & Drake Plastics                                            \\
        DS\label{acr:ds}         & Dielectric Sciences                                       \\
        DUNE\label{acr:dune}     & Deep Underground Neutrino Experiment                      \\
		EC\label{acr:ec}         & Electro Static Dissipative                                \\
		ESC\label{acr:esc}       & Electrostatic Collection                                  \\
		ESD\label{acr:esd}       & Electrically Conductive                                   \\
		FBK\label{acr:fbk}       & Fondazione Bruno Kessler                                  \\
		FE\label{acr:fe}         & Frontend Electronics                                      \\
		FFKM\label{acr:ffkm}     & Perfluoroelastomer                                        \\
		FKM\label{acr:fkm}       & Fluoroelastomer (Viton)                                   \\
		GD-MS\label{acr:gdms}    & Glow Discharge Mass Spectrometry                          \\
		HSSC\label{acr:hssc}     & High Stability Silicon Capacitors                         \\
		HV\label{acr:hv}         & High Voltage                                              \\
		ICP-MS\label{acr:icpms}  & Inductively Coupled Plasma Mass Spectrometry              \\
		IHEP\label{acr:ihep}     & Institute of High Energy Physics                          \\
		IME\label{acr:ime}       & Institute of Microelectronics                             \\
		IP\label{acr:ip}         & Interstate Plastics                                       \\
%                IMS & Institut f\"{u}r Mikroelektronische Schaltungen und Systeme \\
%		ITEP\label{acr:itep}     & Institute for Theoretical and Experimental Physics        \\
		ITO\label{acr:ito}       & Indium Tin Oxide                                          \\
		IZM\label{acr:izm}       & Institut f\"{u}r Zuverl\"{a}ssigkeit und Mikrointegration \\
		LIGO\label{acr:ligo}     & Laser Interferometer Gravitational-Wave Observatory       \\
		LNGS\label{acr:lngs}     & Laboratori Nazionali del Gran Sasso                       \\
		LU\label{acr:lu}         & Laurentian University                                     \\
        LZ\label{acr:lz}         & LUX-ZEPLIN Experiment                                     \\
		MBNL\label{acr:mbnl}     & Marvell Berkeley Nanofabrication Laboratory               \\
		MCAM\label{acr:mcam}     & Mitsubishi Chemical Advanced Materials                    \\
		MIG\label{acr:mig}       & Metal Inert Gas (Welding)                                 \\
		MOSIS\label{acr:mosis}   & Metal Oxide Semiconductor Implementation Service          \\
		MPIK\label{acr:mpik}     & Max-Planck-Institut f\"{u}r Kernphysik                    \\
		MPPC\label{acr:mppc}     & Multi-Pixel Photon Counters                               \\
		NAA\label{acr:naa}       & Neutron Activation Analysis                               \\
		NC\label{acr:nc}         & Nonconducting (defective conductor)                       \\
%                NIST & National Institute of Standards and Technology \\
%                NMC & Nippon Micrometal Corporation \\
		NRC\label{acr:nrc}       & National Research Council Canada                         \\
		OFHC\label{acr:ofhc}     & Oxygen-free High-conductivity                             \\
		OFRP\label{acr:ofrp}     & Oxygen-free radiopure~\cite{Szücs2019}                    \\
		PAI\label{acr:pai}       & Polyamide-imide                                           \\
		PCB\label{acr:pcb}       & Printed Circuit Board                                     \\
        PCTFE\label{acr:pctfe}   & Polychlorotrifluoroethylene                               \\
		PECVD\label{acr:pecvd}   & Plasma-enhanced CVD                                       \\
		PE\label{acr:pe}         & Polyethylene                                              \\
		PEI\label{acr:pei}       & Polyetherimide                                            \\
		PI\label{acr:pi}         & Polyimide                                                 \\
		PMT\label{acr:pmt}       & Patz Materials and Technologies                           \\
		PNNL\label{acr:pnnl}     & Pacific Northwest National Laboratory                     \\
		PP\label{acr:pp}         & Professional Plastics                                     \\
		PTFE\label{acr:ptfe}     & Polytetrafluoroethylene (Teflon)                          \\
        QEP\label{acr:qep}       & Quadrant Engineering Plastic                              \\
        RMD\label{acr:rmd}       & Radiation Monitoring Devices                              \\
		SAES\label{acr:saes}     & Societ\`{a} Apparecchi Elettrici e Scientifici            \\
		SC\label{acr:sc}         & Single Crystal                                            \\
		SCC\label{acr:scc}       & Spencer Composites Corporation                            \\
		SCS\label{acr:scs}       & Southern Copper and Supply                                \\
%                SEH & Shin-Etsu Handotai \\
		SIOM\label{acr:siom}     & Shanghai Institute of Optics and Fine Mechanics           \\
		SiPM\label{acr:sipm}     & Silicon Photomultipliers                                   \\
		SLAC\label{acr:slac}     & SLAC National Accelerator Laboratory                      \\
		SNF\label{acr:snf}       & Stanford Nanofabrication Facility                         \\
%                SPAD & Single Photon Avalanche Detector \\
		SURF\label{acr:surf}     & Sanford Underground Research Facility                    \\
%                TECAPEI: trademarked name
		TIG\label{acr:tig}       & Tungsten Inert Gas (Welding)                              \\
		TMC\label{acr:tmc}       & Technical Marketing Company                               \\
		TSMC\label{acr:tsmc}     & Taiwan Semiconductor Manufacturing Company                \\
		TUNL\label{acr:tunl}     & Triangle University Nuclear Laboratory                    \\
		UA\label{acr:ua}         & University of Alabama                                     \\
		U.G.~Ge\label{acr:ugge}  & Underground Germanium Counting                            \\
		UHMW\label{acr:uhmw}     & Ultra-high-molecular-weight                               \\
		UK\label{acr:uk}         & University of Kentucky                                    \\
		USeoul\label{acr:useoul} & University of Seoul                                       \\
		UV\label{acr:uv}         & Ultraviolet                                               \\
		VCSEL\label{acr:vcsel}   & Vertical-cavity surface-emitting Laser                    \\
		VdA\label{acr:vda}       & Vue-des-Alpes detector operated by the University of Bern and later SNOLAB \\
%                VUV & Vacuum Ultraviolet%
                 \end{tblr}%
        \end{ruledtabular}%
    \end{table}%
\endgroup%

%% file: bigtable2.tex
%
% Setup sample count
\newcounter{sampleno}%
\DeclareRobustCommand{\sample}[1]{%
   \refstepcounter{sampleno}%
   \thesampleno\label{#1}%
}%
\crefalias{sampleno}{sample}%
%
%
%
% Setup material column width; a maximum given the widths required for the other columns
\newlength{\materialwidth}%
\setlength{\materialwidth}{268pt}% for columns with aligned decimal places
\NewColumnType{M}{Q[m,\materialwidth]}% cannot use calc or variable lengths in tblr so set it up here
%
%\begingroup%
\squeezetable%
\SetTblrInner{rowsep=0pt}
\UseTblrLibrary{counter}
\begin{turnpage}%
    \begin{table*}[t]%
        \caption{\label{tab:bigtable}Table of radioassay. Additional isotopic data is listed in table \ref{tab:bigtablex} for entries marked with $^\dagger$.}%
        \begin{tblr}{width=\linewidth,colsep=2pt,colspec={%
                @{}r|%
                M%
                @{}l%
                l%
                @{}Q[c, si={table-align-comparator=false,table-format=<4.2+-1.2e1}]%
                @{}Q[c, si={table-align-comparator=false,table-format=<4.2+-1.2e1}]%
                @{}Q[c, si={table-align-comparator=false,table-format=<4.2+-1.2e1}]%
                @{}Q[c, si={table-align-comparator=false,table-format=<4.2+-1.2e1}]%
            }%
        }%
        \hline\hline%
        \# & Material & Technique & Institute & {U early [ppt]}  & {U late [ppt]}  & {Th [ppt]} & {K [ppb]}\\
\hline%
& \textbf{Copper}\\
%BS EN 197
\sample{R-045.1.1}$^\dagger$             & \hyperref[acr:scs]{SCS} \hyperref[acr:ofhc]{OFHC} C10100%
&\hyperref[acr:agge]{A.G.\,Ge} & \hyperref[acr:ua]{UA}
&                          & <20                      & <43                      & <40                      \\
\sample{R-045.1.2}\phantom{$^\dagger$}   & \hyperref[acr:scs]{SCS} \hyperref[acr:ofhc]{OFHC} C10100%
&\hyperref[acr:gdms]{GD-MS}    & \hyperref[acr:nrc]{NRC}
& <20                      &                           & <20                      & <0.70                    \\
\sample{R-045.1.3}$^\dagger$             & \hyperref[acr:scs]{SCS} \hyperref[acr:ofhc]{OFHC} C10100%
&\hyperref[acr:agge]{A.G.\,Ge} & \hyperref[acr:ua]{UA}
&                          & <8.7                     & <28                      & <32                      \\
\sample{R-072.1.1}\phantom{$^\dagger$}   & Canberra Ge crystal holder%
&\hyperref[acr:gdms]{GD-MS}    & \hyperref[acr:nrc]{NRC}
& <30                      &                           & <30                      & <0.60                    \\
\sample{R-080.1.1}\phantom{$^\dagger$}   & Norddeutsche Affinerie \hyperref[acr:ofrp]{OFRP}, EXO-200 cryostat%
&\hyperref[acr:gdms]{GD-MS}    & \hyperref[acr:nrc]{NRC}
& <20                      &                           & <20                      & <0.60                    \\
\sample{R-136.1.1}\phantom{$^\dagger$}   & GRIKIN Advanced Materials, target for \hyperref[acr:pecvd]{PECVD}%
&\hyperref[acr:icpms]{ICP-MS}  & \hyperref[acr:ihep]{IHEP}
& <0.34                    &                           & <0.35                    &                           \\
\sample{R-168.1.1}\phantom{$^\dagger$}   & Electroformed by \hyperref[acr:pnnl]{PNNL}%
&\hyperref[acr:icpms]{ICP-MS}  & \hyperref[acr:pnnl]{PNNL}
& <0.0094                  &                           & 0.0060+-0.0010          &                           \\
\cline{2-2}%
& \qquad Aurubis copper\\
\sample{R-002.1.1}\phantom{$^\dagger$}   & Anode before electrolysis%
&\hyperref[acr:gdms]{GD-MS}    & \hyperref[acr:nrc]{NRC}
& <20                      &                           & <20                      & <0.90                    \\
\sample{R-002.1.2}\phantom{$^\dagger$}   & Anode before electrolysis%
&\hyperref[acr:icpms]{ICP-MS}  & \hyperref[acr:useoul]{USeoul}
& 3.1+-1.1                &                           & 2.4+-0.9                &                           \\
\sample{R-002.2.1}\phantom{$^\dagger$}   & Cathode after electrolysis%
&\hyperref[acr:gdms]{GD-MS}    & \hyperref[acr:nrc]{NRC}
& <20                      &                           & <20                      & <0.90                    \\
\sample{R-002.2.2}\phantom{$^\dagger$}   & Cathode after electrolysis%
&\hyperref[acr:icpms]{ICP-MS}  & \hyperref[acr:useoul]{USeoul}
& <0.84                    &                           & <1.2                     &                           \\
\sample{R-002.3.1}\phantom{$^\dagger$}   & Cast and rolled wire%
&\hyperref[acr:gdms]{GD-MS}    & \hyperref[acr:nrc]{NRC}
& <20                      &                           & <20                      & <0.80                    \\
\sample{R-002.3.2}\phantom{$^\dagger$}   & Cast and rolled wire%
&\hyperref[acr:icpms]{ICP-MS}  & \hyperref[acr:useoul]{USeoul}
& <0.34                    &                           & 1.6+-1.1                &                           \\
\sample{R-002.4.1}\phantom{$^\dagger$}   & Electrolytic cathode (ASTM B115 and BS EN 1978)%
&\hyperref[acr:gdms]{GD-MS}    & \hyperref[acr:nrc]{NRC}
& <30                      &                           & <30                      & <0.80                    \\
\sample{R-002.5.1}\phantom{$^\dagger$}   & Electrolytic cathode (ASTM B115 and BS EN 1978)%
&\hyperref[acr:gdms]{GD-MS}    & \hyperref[acr:nrc]{NRC}
& <30                      &                           & <30                      & <0.80                    \\
\sample{R-002.6.1}\phantom{$^\dagger$}   & Electrolytic cathode (ASTM B115 and BS EN 1978)%
&\hyperref[acr:gdms]{GD-MS}    & \hyperref[acr:nrc]{NRC}
& <30                      &                           & <30                      & <1.0                     \\
\sample{R-002.7.1}$^\dagger$             & Anode%
&\hyperref[acr:ugge]{U.G.~Ge}  & \hyperref[acr:vda]{VdA}
&                          & 15.8+-1.1               & 27.8+-2.2               & 3.7+-1.9                \\
\sample{R-002.7.2}$^\dagger$             & Anode%
&\hyperref[acr:ugge]{U.G.~Ge}  & \hyperref[acr:vda]{VdA}
&                          & <0.40                    & 6.9+-2.6                & <7.4                     \\
\sample{R-002.7.3}\phantom{$^\dagger$}   & Anode%
&\hyperref[acr:icpms]{ICP-MS}  & \hyperref[acr:cup]{CUP}
& <16                      &                           & 104+-9                  &                           \\
\sample{R-002.7.4}\phantom{$^\dagger$}   & Anode%
&\hyperref[acr:icpms]{ICP-MS}  & \hyperref[acr:cup]{CUP}
& <9.6                     &                           & <21                      &                           \\
\sample{R-002.7.5}\phantom{$^\dagger$}   & Anode%
&\hyperref[acr:icpms]{ICP-MS}  & \hyperref[acr:cup]{CUP}
& <11                      &                           & <20                      &                           \\
\sample{R-002.8.1}$^\dagger$             & Cathode%
&\hyperref[acr:ugge]{U.G.~Ge}  & \hyperref[acr:vda]{VdA}
&                          & <1.2                     & <2.2                     & 3.0+-2.0                \\
\sample{R-002.8.2}\phantom{$^\dagger$}   & Cathode%
&\hyperref[acr:icpms]{ICP-MS}  & \hyperref[acr:cup]{CUP}
& <7.1                     &                           & <11                      &                           \\
\sample{R-002.11.1}\phantom{$^\dagger$}  & Cathode%
&\hyperref[acr:icpms]{ICP-MS}  & \hyperref[acr:pnnl]{PNNL}
& 0.254+-0.008            &                           & 0.13+-0.06              &                           \\
\sample{R-002.12.1}\phantom{$^\dagger$}  & Cathode%
&\hyperref[acr:gdms]{GD-MS}    & \hyperref[acr:nrc]{NRC}
& <20                      &                           & <20                      & <0.70                    \\
\hline%
& \textbf{Titanium and Nickel}\\
\sample{R-005.1.2}\phantom{$^\dagger$}   & VSMPO-AVISMA titanium (VT1-00)% \hyperref[acr:itep]{ITEP}%
&\hyperref[acr:gdms]{GD-MS}    & \hyperref[acr:nrc]{NRC}
& 1000(1000:500)      &                           & <80                      & <4.0                     \\
\sample{R-167.1.1}\phantom{$^\dagger$}   & CVMR \hyperref[acr:cvd]{CVD} nickel, raw dilution%
&\hyperref[acr:icpms]{ICP-MS}  & \hyperref[acr:pnnl]{PNNL}
& <1.6                     &                           & <0.99                    &                           \\
\sample{R-167.1.2}\phantom{$^\dagger$}   & CVMR \hyperref[acr:cvd]{CVD} nickel, after matrix separation%
&\hyperref[acr:icpms]{ICP-MS}  & \hyperref[acr:pnnl]{PNNL}
& 0.040+-0.016            &                           & 0.80+-0.30              &                           \\
\sample{R-207.1.1}\phantom{$^\dagger$}   & Western Grade CVMR \hyperref[acr:cvd]{CVD} nickel \cite{ROOSENDAAL2026171402}%
&\hyperref[acr:icpms]{ICP-MS}  & \hyperref[acr:pnnl]{PNNL}
& <0.098                   &                           & 0.065+-0.029            & 0.91+-0.10              \\
\hline%
& \textbf{Aluminum}\\
\sample{R-057.1.1}\phantom{$^\dagger$}   & Canberra high purity%
&\hyperref[acr:gdms]{GD-MS}    & \hyperref[acr:nrc]{NRC}
& <100                     &                           & <100                     & <5.0                     \\
\sample{R-141.1.1}$^\dagger$             & Materion Advanced Chemicals pellets, 99.999\% pure%
&\hyperref[acr:ugge]{U.G.~Ge}  & \hyperref[acr:surf]{SURF}%
& 70+-40                  & <410                     & 490+-120                & 130+-90                 \\
\sample{R-160.1.1}\phantom{$^\dagger$}   & Norsk Hydro%
&\hyperref[acr:gdms]{GD-MS}    & \hyperref[acr:nrc]{NRC}
& <70                      &                           & <70                      & <5.0                     \\
\sample{R-161.1.1}\phantom{$^\dagger$}   & Norsk Hydro%
&\hyperref[acr:gdms]{GD-MS}    & \hyperref[acr:nrc]{NRC}
& <60                      &                           & <70                      & <4.0                     \\
\sample{R-162.1.1}\phantom{$^\dagger$}   & Norsk Hydro%
&\hyperref[acr:gdms]{GD-MS}    & \hyperref[acr:nrc]{NRC}
& <80                      &                           & <90                      & <5.0                     \\
\sample{R-166.1.1}\phantom{$^\dagger$}   & Norsk Hydro%
&\hyperref[acr:icpms]{ICP-MS}  & \hyperref[acr:pnnl]{PNNL}
& 145+-9                 &                           &  134+-11                 &                           \\
\sample{R-166.2.1}\phantom{$^\dagger$}   & Norsk Hydro%
&\hyperref[acr:icpms]{ICP-MS}  & \hyperref[acr:pnnl]{PNNL}
& 160+-60                 &                           &  <430                     &                           \\
\sample{R-166.3.1}\phantom{$^\dagger$}   & Norsk Hydro%
&\hyperref[acr:icpms]{ICP-MS}  & \hyperref[acr:pnnl]{PNNL}
& 145.0+-2.1              &                           &  366+-32                 &                           \\
\hline%
& \textbf{Silicon}\\
\sample{R-085.1.1}\phantom{$^\dagger$}   & El-Cat \hyperref[acr:sc]{SC} wafer, high resistivity float zone (D161)%
&\hyperref[acr:naa]{NAA}       & \hyperref[acr:ua]{UA}
& <0.63                    &                           &  <0.33                    &  0.50+-0.06              \\
\sample{R-124.1.1}\phantom{$^\dagger$}   & Topsil \hyperref[acr:sc]{SC}, float zone, 33-1204-10 seed end%
&\hyperref[acr:icpms]{ICP-MS}  & \hyperref[acr:pnnl]{PNNL}
& 0.74+-0.26              &                           &  3.4+-1.1                &  2.6+-0.7                \\
\sample{R-124.2.1}\phantom{$^\dagger$}   & Topsil \hyperref[acr:sc]{SC}, float zone, 33-1204-40 central%
&\hyperref[acr:icpms]{ICP-MS}  & \hyperref[acr:pnnl]{PNNL}
& 1.0+-0.4                &                           &  2.6+-0.6                &  2.0+-0.7                \\
\sample{R-081.1.1}\phantom{$^\dagger$}   & Topsil \hyperref[acr:sc]{SC}, float zone, 33-1204-40 central, machined by McCarter Machine%
&\hyperref[acr:naa]{NAA}       & \hyperref[acr:ua]{UA}
& <1.8                     &                           &  <0.16                    &  7.1+-0.7                \\
\sample{R-124.4.1}\phantom{$^\dagger$}   & Topsil \hyperref[acr:sc]{SC}, float zone, 33-1204-40 non-seed end%
&\hyperref[acr:icpms]{ICP-MS}  & \hyperref[acr:pnnl]{PNNL}
& 1.12+-0.15              &                           &  3.3+-0.4                &  3.6+-1.2                \\
        \end{tblr}%
    \end{table*}%
\end{turnpage}%
\begin{turnpage}%
    \begin{table*}[t]%
        \begin{tblr}{width=\linewidth,colsep=2pt,colspec={%
                @{}r|%
                M%
                @{}l%
                l%
                @{}Q[c, si={table-align-comparator=false,table-format=<4.2+-1.2e1}]%
                @{}Q[c, si={table-align-comparator=false,table-format=<4.2+-1.2e1}]%
                @{}Q[c, si={table-align-comparator=false,table-format=<4.2+-1.2e1}]%
                @{}Q[c, si={table-align-comparator=false,table-format=<4.2+-1.2e1}]%
                }%
        }%
\hline%
& \textbf{Sapphire} \\
\sample{R-046.1.1}\phantom{$^\dagger$}   & GT Advanced Technology/\hyperref[acr:ligo]{LIGO}%
&\hyperref[acr:naa]{NAA}       & \hyperref[acr:ua]{UA}
& <1.8                     &                           &  6.0+-1.2                &  9.5+-2.2                \\
\sample{R-075.1.1}$^\dagger$             & GT Advanced Technology crackle%
&\hyperref[acr:agge]{A.G.\,Ge} & \hyperref[acr:ua]{UA}
&                          &  <270                     &  <1100                    &  1800+-1200              \\
\sample{R-048.1.1}\phantom{$^\dagger$}   & Saint-Gobain Crystals, as-grown%
&\hyperref[acr:naa]{NAA}       & \hyperref[acr:ua]{UA}
& 2.6+-0.4e4      &                           &                           &  8.1+-2.8                \\
\sample{R-048.6.1}\phantom{$^\dagger$}   & Saint-Gobain Crystals, ground, cut, annealed%
&\hyperref[acr:naa]{NAA}       & \hyperref[acr:ua]{UA}
& 5.1+-3.8                &                           &  <0.10                   &  3.0+-1.1                \\
\sample{R-048.6.2}\phantom{$^\dagger$}   & Saint-Gobain Crystals, ground, cut, annealed \cite{zgzm-w5f4}%
&\hyperref[acr:naa]{NAA}       & \hyperref[acr:tunl]{TUNL}
& <1.4                     &                           &  0.53+-0.32              &                           \\
\sample{R-056.1.1}$^\dagger$             & Saint-Gobain  Crystals alumina (SA27628, lot 1605)%
&\hyperref[acr:agge]{A.G.\,Ge} & \hyperref[acr:ua]{UA}
&                          &  <790                     &  3.3+-0.4e4      &  6100+-1500              \\
\sample{R-056.2.1}$^\dagger$             & Saint-Gobain  Crystals alumina (SA19645, lot 1555.2)%
&\hyperref[acr:agge]{A.G.\,Ge} & \hyperref[acr:ua]{UA}
&                          &  <270                     &  1800+-600               &  <580                     \\
\sample{R-056.3.1}$^\dagger$             & Saint-Gobain  Crystals alumina (SA27335, lot 1617.2)%
&\hyperref[acr:agge]{A.G.\,Ge} & \hyperref[acr:ua]{UA}
&                          &  2600+-600               &  6500+-1300              &  <1200                    \\
\sample{R-084.1.1}\phantom{$^\dagger$}   & Precision Sapphire Technologies%
&\hyperref[acr:naa]{NAA}       & \hyperref[acr:ua]{UA}
& 990+-100                &                           &  410+-40                 &  3.4+-1.6                \\
\hline%
& \textbf{Fused silica} \\
\sample{R-043.1.1}\phantom{$^\dagger$}   & \hyperref[acr:siom]{SIOM} molten, re-formed natural quartz%
&\hyperref[acr:naa]{NAA}       & \hyperref[acr:ua]{UA}
& 340+-50                 &                           &  11.7+-1.7               &  290+-33                 \\
\sample{R-055.1.1}\phantom{$^\dagger$}   & Corning \hyperref[acr:uv]{UV} grade (7980 1-D)%
&\hyperref[acr:naa]{NAA}       & \hyperref[acr:ua]{UA}
& 550+-50                 &                           &                           &  4.3+-0.5                \\
\sample{R-078.1.1}\phantom{$^\dagger$}   & Heraeus Spectrosil 2000%
&\hyperref[acr:naa]{NAA}       & \hyperref[acr:ua]{UA}
& <1.4                     &                           &  <0.11                    &  0.55+-0.06              \\
%
% Provided by Darkside:
%\sample{R-044.1.1}\phantom{$^\dagger$}   & Heraeus Suprasil 312%
%&\hyperref[acr:naa]{NAA}       & \hyperref[acr:ua]{UA}
%& <4.2                     &                           &  <2.6                     &                           \\
%
\sample{R-192.1.1.1}\phantom{$^\dagger$} & \hyperref[acr:snf]{SNF} Heraeus Suprasil 300%
&\hyperref[acr:icpms]{ICP-MS}  & \hyperref[acr:pnnl]{PNNL}
& 23.6+-0.6               &                           &  4.40+-0.20              &  3.09+-0.15              \\
\sample{R-192.1.1.2}\phantom{$^\dagger$} & \hyperref[acr:snf]{SNF} Heraeus Suprasil 300, Si \hyperref[acr:cvd]{CVD} @ \qty{530}{\degreeCelsius}, Ti and Pd end metalized%
&\hyperref[acr:icpms]{ICP-MS}  & \hyperref[acr:pnnl]{PNNL}
& 5.41+-0.10              &                           &  7.1+-0.5                &  13.00+-0.20             \\
\sample{R-192.1.1.3}\phantom{$^\dagger$} & \hyperref[acr:snf]{SNF} Heraeus Suprasil 300, Si \hyperref[acr:cvd]{CVD} @ \qty{580}{\degreeCelsius}, Ti and Pd end metalized%
&\hyperref[acr:icpms]{ICP-MS}  & \hyperref[acr:pnnl]{PNNL}
& 10.6+-0.4               &                           &  7.60+-0.30              &  78.8+-1.1               \\
\sample{R-219.1.1.1}\phantom{$^\dagger$} & \hyperref[acr:mbnl]{MBNL} Heraeus Suprasil 300%
&\hyperref[acr:icpms]{ICP-MS}  & \hyperref[acr:pnnl]{PNNL}
& 0.220+-0.020            &                           &  0.72+-0.09              &  <0.63                    \\
\sample{R-219.1.1.2}\phantom{$^\dagger$} & \hyperref[acr:mbnl]{MBNL} Heraeus Suprasil 300, aSi \hyperref[acr:cvd]{CVD} @ \qty{550}{\degreeCelsius}, no metalization%
&\hyperref[acr:icpms]{ICP-MS}  & \hyperref[acr:pnnl]{PNNL}
& 0.540+-0.030            &                           &  0.82+-0.06              &  15.40+-0.20             \\
\sample{R-219.1.1.3}\phantom{$^\dagger$} & \hyperref[acr:mbnl]{MBNL} Heraeus Suprasil 300, aSi \hyperref[acr:cvd]{CVD} @ \qty{550}{\degreeCelsius}, Ti and Pd end metalized%
&\hyperref[acr:icpms]{ICP-MS}  & \hyperref[acr:pnnl]{PNNL}
& 4.30+-0.20              &                           &  2.00+-0.16              &  7.90+-0.20              \\
\hline%
& \textbf{Carbon fiber composite} \\
\sample{R-016.1.1}$^\dagger$             & \hyperref[acr:scc]{SCC} resin, cured-ESM611+%
&\hyperref[acr:agge]{A.G.\,Ge} & \hyperref[acr:ua]{UA}
&                          &  <350                     &  <2800                    &  <6000                    \\
\sample{R-016.1.2}$^\dagger$             & \hyperref[acr:scc]{SCC} resin, cured-ESM611+%
&\hyperref[acr:agge]{A.G.\,Ge} & \hyperref[acr:ua]{UA}
&                          &  <160                     &  450+-230                &  <990                     \\
\sample{R-016.1.3}$^\dagger$             & \hyperref[acr:scc]{SCC} resin, cured-ESM611+%
&\hyperref[acr:ugge]{U.G.~Ge}  & SNOLAB
&                          &  100+-40                 &  <73                      &  <320                     \\
\sample{R-016.2.1}$^\dagger$             & \hyperref[acr:scc]{SCC} resin, cured-ESM70%
&\hyperref[acr:agge]{A.G.\,Ge} & \hyperref[acr:ua]{UA}
&                          &  <1700                    &  <1400                    &  <4200                    \\
\sample{R-016.2.2}$^\dagger$             & \hyperref[acr:scc]{SCC} resin, cured-ESM70%
&\hyperref[acr:agge]{A.G.\,Ge} & \hyperref[acr:ua]{UA}
&                          &  <130                     &  <430                     &  <220                     \\
\sample{R-016.2.3}$^\dagger$             & \hyperref[acr:scc]{SCC} resin, cured-ESM70%
&\hyperref[acr:ugge]{U.G.~Ge}  & SNOLAB
&                          &  <73                      &  740+-220                &  <170                     \\
\sample{R-016.3.1}$^\dagger$             & \hyperref[acr:scc]{SCC} resin, cured-862/81k%
&\hyperref[acr:agge]{A.G.\,Ge} & \hyperref[acr:ua]{UA}
&                          &  <720                     &  <1700                    &  <9400                    \\
\sample{R-016.3.2}$^\dagger$             & \hyperref[acr:scc]{SCC} resin, cured-862/81k%
&\hyperref[acr:agge]{A.G.\,Ge} & \hyperref[acr:ua]{UA}
&                          &  <300                     &  <240                     &  <100                     \\
\sample{R-016.3.3}$^\dagger$             & \hyperref[acr:scc]{SCC} resin, cured-862/81k%
&\hyperref[acr:ugge]{U.G.~Ge}  & SNOLAB
&                          &  <34                      &  <200                     &  <160                     \\
\sample{R-016.3.4}$^\dagger$             & \hyperref[acr:scc]{SCC} resin, cured-862/81k%
&\hyperref[acr:ugge]{U.G.~Ge}  & \hyperref[acr:vda]{VdA}
&                          &  <6.0                     &  <8.3                     &  <21                      \\
\sample{R-017.1.1}$^\dagger$             & Mitsubishi Rayon Pyrofil fiber%
&\hyperref[acr:agge]{A.G.\,Ge} & \hyperref[acr:ua]{UA}
&                          &  <350                     &  <890                     &  <5100                    \\
\sample{R-017.1.2}$^\dagger$             & Mitsubishi Rayon Pyrofil fiber%
&\hyperref[acr:ugge]{U.G.~Ge}  & SNOLAB
&                          &  <29                      &  <250                     &  510+-200                \\
\sample{R-017.1.3}$^\dagger$             & Mitsubishi Rayon Pyrofil fiber%
&\hyperref[acr:ugge]{U.G.~Ge}  & \hyperref[acr:vda]{VdA}
&                          &  82+-34                  &  280+-90                 &  670+-170                \\
\sample{R-017.2.1}$^\dagger$             & Grafil fiber (34-12K, lot 2700I)%
&\hyperref[acr:agge]{A.G.\,Ge} & \hyperref[acr:ua]{UA}
&                          &  <230                     &  <340                     &  1600+-500               \\
\sample{R-017.2.2}$^\dagger$             & Grafil fiber (34-12K, lot 2700I)%
&\hyperref[acr:ugge]{U.G.~Ge}  & SNOLAB
&                          &  104+-23                 &  <240                     &  2150+-220               \\
\sample{R-017.2.3}$^\dagger$             & Grafil fiber (34-12K, lot 2700I)%
&\hyperref[acr:ugge]{U.G.~Ge}  & \hyperref[acr:vda]{VdA}
&                          &  40+-14                  &  130+-40                 &  1080+-100               \\
\sample{R-017.3.1}$^\dagger$             & Grafil fiber (34-12K R, lot 2922I)%
&\hyperref[acr:agge]{A.G.\,Ge} & \hyperref[acr:ua]{UA}
&                          &  <140                     &  420+-310                &  1780+-280               \\
\sample{R-017.3.2}$^\dagger$             & Grafil fiber (34-12K R, lot 2922I)%
&\hyperref[acr:ugge]{U.G.~Ge}  & SNOLAB
&                          &  <110                     &  <270                     &  1130+-270               \\
\sample{R-017.3.3}$^\dagger$             & Grafil fiber (34-12K R, lot 2922I)%
&\hyperref[acr:ugge]{U.G.~Ge}  & \hyperref[acr:vda]{VdA}
&                          &  41+-15                  &  70+-40                  &  790+-90                 \\
\sample{R-039.1.1}$^\dagger$             & EPON 862-W block%
&\hyperref[acr:ugge]{U.G.~Ge}  & SNOLAB
&                          &  <25                      &  <860                     &  <410                     \\
\sample{R-041.1.1}$^\dagger$             & Hexcel IM-7-GP-6k fiber%
&\hyperref[acr:agge]{A.G.\,Ge} & \hyperref[acr:ua]{UA}
&                          &  <5700                    &  4.7+-2.0e4              &  <3.2e4                   \\
\sample{R-038.1.1}$^\dagger$             & Hexcel IM-7 fiber + \hyperref[acr:pmt]{PMT}-F1 resin, plate%
&\hyperref[acr:ugge]{U.G.~Ge}  & SNOLAB
&                          &  <550                     &  <660                     &  2700+-1100              \\
\sample{R-037.1.1}$^\dagger$             & Tenax HTS40-12k fiber + EPON 862-W resin, tube%
&\hyperref[acr:ugge]{U.G.~Ge}  & SNOLAB
&                          &  <990                     &  <4400                    &  1.9+-0.5e4              \\
\sample{R-036.1.1}$^\dagger$             & Toray T1000-12k fiber + EPON 862-W-BF3 resin, tube%
&\hyperref[acr:ugge]{U.G.~Ge}  & SNOLAB
&                          &  <140                     &  <6400                    &  4900+-3200              \\
\sample{R-054.2.1}$^\dagger$             & Toray fiber%
&\hyperref[acr:ugge]{U.G.~Ge}  & SNOLAB
&                          &  <7.4                     &  <250                     &  1920+-250               \\
\sample{R-054.1.1}$^\dagger$             & Zoltek (PX30MF0150, lot 1M14015)%
&\hyperref[acr:ugge]{U.G.~Ge}  & SNOLAB
&                          &  1390+-150               &  1800+-400               &  1.26+-0.13e4            \\
              \end{tblr}%
    \end{table*}%
\end{turnpage}%
\begin{turnpage}%
    \begin{table*}[t]%
        \begin{tblr}{width=\linewidth,colsep=2pt,colspec={%
                @{}r|%
                M%
                @{}l%
                l%
                @{}Q[c, si={table-align-comparator=false,table-format=<4.2+-1.2e1}]%
                @{}Q[c, si={table-align-comparator=false,table-format=<4.2+-1.2e1}]%
                @{}Q[c, si={table-align-comparator=false,table-format=<4.2+-1.2e1}]%
                @{}Q[c, si={table-align-comparator=false,table-format=<4.2+-1.2e1}]%
            }%
        }%
\hline%
& \textbf{Viton} \\
\sample{R-062.1.1}$^\dagger$             & Canberra o-rings%
&\hyperref[acr:agge]{A.G.\,Ge} & \hyperref[acr:ua]{UA}
&                          &  6.6+-0.8e4              &  1.7+-0.7e4              &  <2.0e4                   \\
\sample{R-112.1.1}\phantom{$^\dagger$}   & Marco Rubber white FKM o-ring (V1012-154)%
&\hyperref[acr:agge]{A.G.\,Ge} & \hyperref[acr:ua]{UA}
&                          &  7.0+-0.9e4              &  1.37+-0.19e5            &  9.2+-2.4e4              \\
\sample{R-113.1.1}\phantom{$^\dagger$}   & Marco Rubber off white FKM o-ring (\hyperref[acr:pai]{PAI}-9725-151)%
&\hyperref[acr:agge]{A.G.\,Ge} & \hyperref[acr:ua]{UA}
&                          &  2.33+-0.24e5            &  4.4+-0.5e4              &  2.19+-0.24e5            \\
\sample{R-114.1.1}\phantom{$^\dagger$}   & Marco Rubber dark brown FKM o-ring (16110247)%
&\hyperref[acr:agge]{A.G.\,Ge} & \hyperref[acr:ua]{UA}
&                          &  2.41+-0.24e5            &  2.75+-0.27e5            &  8.4+-1.4e4              \\
\sample{R-115.1.1}\phantom{$^\dagger$}   & Marco Rubber light brown FKM o-ring (EV17550304)%
&\hyperref[acr:agge]{A.G.\,Ge} & \hyperref[acr:ua]{UA}
&                          &  2.09+-0.21e5            &  1.01+-0.10e5            &  8.5+-0.9e4              \\
\sample{R-111.1.1}\phantom{$^\dagger$}   & Marco Rubber off white FKM o-ring (TBD-152)%
&\hyperref[acr:agge]{A.G.\,Ge} & \hyperref[acr:ua]{UA}
&                          &  7.3+-0.7e4              &  1.17+-0.17e5            &  6.5+-0.6e4              \\
\sample{R-123.1.1}\phantom{$^\dagger$}   & Marco Rubber white FKM o-ring (TBD-152)%
&\hyperref[acr:agge]{A.G.\,Ge} & \hyperref[acr:ua]{UA}
&                          &  5.1+-0.5e4              &  3.33+-0.34e4            &  1.15+-0.24e5            \\
\sample{R-074.2.1}$^\dagger$             & Marco Rubber white FFKM o-rings (Z1216-152)%
&\hyperref[acr:agge]{A.G.\,Ge} & \hyperref[acr:ua]{UA}
&                          &  <1400                    &  7000+-4000              &  <2.3e4                   \\
\sample{R-074.1.1}$^\dagger$             & Marco Rubber black FFKM  o-rings (Z1319-152)%
&\hyperref[acr:agge]{A.G.\,Ge} & \hyperref[acr:ua]{UA}
&                          &  9.5+-1.1e4              &  2.34+-0.26e5            &  1.34+-0.22e5            \\
\hline%
& \textbf{Acrylic} \\
\sample{R-030.1.1}\phantom{$^\dagger$}   & Shanghai Red Coral%
&\hyperref[acr:agge]{A.G.\,Ge} & \hyperref[acr:ua]{UA}
&                          &  <120                     &  310+-180                &  <180                     \\
\sample{R-030.2.1}$^\dagger$             & Jiangyin Golden Harvest%
&\hyperref[acr:agge]{A.G.\,Ge} & \hyperref[acr:ua]{UA}
&                          &  100+-50                 &  <45                      &  <91                      \\
\sample{R-100.1.1}\phantom{$^\dagger$}   & SciCron \hyperref[acr:ito]{ITO}-coated (AC-300)%
&\hyperref[acr:icpms]{ICP-MS}  & \hyperref[acr:pnnl]{PNNL}
& 6.7+-1.0                &                           &  7.0+-0.7                &                           \\
\hline%
& \textbf{Conductive polymer}\\
\sample{R-021.1.1}$^\dagger$             & \hyperref[acr:ds]{DS} semi-conductive-\hyperref[acr:pe]{PE} \hyperref[acr:hv]{HV} cable 2308%
&\hyperref[acr:ugge]{U.G.~Ge}  & \hyperref[acr:vda]{VdA}
&                          &  500+-260                &  <970                     &  <1700                    \\
\sample{R-021.1.2}$^\dagger$             & \hyperref[acr:ds]{DS} semi-conductive-\hyperref[acr:pe]{PE} \hyperref[acr:hv]{HV} cable 2308%
&\hyperref[acr:ugge]{U.G.~Ge}  & \hyperref[acr:vda]{VdA}
&                          &  <180                     &  <54                      &  <750                     \\
\sample{R-021.2.1}$^\dagger$             & \hyperref[acr:ds]{DS} semi-conductive-\hyperref[acr:pe]{PE} \hyperref[acr:hv]{HV} cable 2343, layers A and B%
&\hyperref[acr:ugge]{U.G.~Ge}  & SNOLAB
&                          &  <170                     &  1630+-290               &  <640                     \\
\sample{R-021.3.1}$^\dagger$             & \hyperref[acr:ds]{DS} semi-conductive-\hyperref[acr:pe]{PE} \hyperref[acr:hv]{HV} cable 2343, layers A--C%
&\hyperref[acr:ugge]{U.G.~Ge}  & SNOLAB
&                          &  490+-70                 &  <380                     &  770+-280                \\
\sample{R-021.8.1}$^\dagger$             & \hyperref[acr:ds]{DS} semi-conductive-\hyperref[acr:pe]{PE} \hyperref[acr:hv]{HV} cable 2343, layers A--C, cleaned%
&\hyperref[acr:ugge]{U.G.~Ge}  & SNOLAB
&                          &  100+-50                 &  <340                     &  610+-240                \\
\sample{R-021.4.1}\phantom{$^\dagger$}   & \hyperref[acr:ds]{DS} semi-conductive-\hyperref[acr:pe]{PE} \hyperref[acr:hv]{HV} cable 2353 (2014 batch), layers A--C%
&\hyperref[acr:naa]{NAA}       & \hyperref[acr:ua]{UA}
& 100+-40                 &                           &  64+-7                   &  350+-40                 \\
\sample{R-021.5.1}\phantom{$^\dagger$}   & \hyperref[acr:ds]{DS} semi-conductive-\hyperref[acr:pe]{PE} \hyperref[acr:hv]{HV} cable 2353 (2014 batch), layer A%
&\hyperref[acr:naa]{NAA}       & \hyperref[acr:ua]{UA}
& 440+-200                &                           &  370+-40                 &  1990+-220               \\
\sample{R-021.6.1}\phantom{$^\dagger$}   & \hyperref[acr:ds]{DS} semi-conductive-\hyperref[acr:pe]{PE} \hyperref[acr:hv]{HV} cable 2353 (2014 batch), layer B%
&\hyperref[acr:naa]{NAA}       & \hyperref[acr:ua]{UA}
& <110                     &                           &  18.6+-2.0               &  132+-14                 \\
\sample{R-021.7.1}\phantom{$^\dagger$}   & \hyperref[acr:ds]{DS} semi-conductive-\hyperref[acr:pe]{PE} \hyperref[acr:hv]{HV} cable 2353 (2014 batch), layer C%
&\hyperref[acr:naa]{NAA}       & \hyperref[acr:ua]{UA}
& 760+-330                &                           &  900+-100                &  2690+-300               \\
\sample{R-083.1.1}$^\dagger$             & \hyperref[acr:ds]{DS} semi-conductive-\hyperref[acr:pe]{PE} \hyperref[acr:hv]{HV} cable 2353 (2014 batch), layers A--C%
&\hyperref[acr:ugge]{U.G.~Ge}  & SNOLAB
&                          &  <160                     &  <500                     &  <990                     \\
\sample{R-206.2.1}$^\dagger$             & \hyperref[acr:ds]{DS} semi-conductive-\hyperref[acr:pe]{PE} \hyperref[acr:hv]{HV} cable 2353 (2023 batch), layers A--C%
&\hyperref[acr:agge]{A.G.\,Ge} & \hyperref[acr:ua]{UA}
& <1.7e4                   &  <580                     &  900+-500                &  3000+-600               \\
\sample{R-213.3.1}$^\dagger$             & \hyperref[acr:ds]{DS} semi-conductive-\hyperref[acr:pe]{PE} \hyperref[acr:hv]{HV} cable 2353, all layers (\hyperref[acr:dune]{DUNE}-\hyperref[acr:nc]{NC})%
&\hyperref[acr:agge]{A.G.\,Ge} & \hyperref[acr:ua]{UA}
& <4300                    &  <330                     &  <1100                    &  2000+-400               \\
\sample{R-101.1.1}\phantom{$^\dagger$}   & McMaster-Carr C-filled \hyperref[acr:ptfe]{PTFE} (8775K56)%
&\hyperref[acr:icpms]{ICP-MS}  & \hyperref[acr:pnnl]{PNNL}
& 11.5+-1.7               &                           &  14.1+-0.6               &                           \\
\sample{R-102.1.1}\phantom{$^\dagger$}   & Applied Plastics Technology C-filled \hyperref[acr:ptfe]{PTFE} (08120891)%
&\hyperref[acr:icpms]{ICP-MS}  & \hyperref[acr:pnnl]{PNNL}
& 14.1+-1.0               &                           &  11.4+-0.4               &                           \\
\sample{R-119.1.1}\phantom{$^\dagger$}   & CS Hyde 1\% C-loaded semi-conductive \hyperref[acr:uhmw]{UHMW} \hyperref[acr:pe]{PE} (19-5F-BLK-24)%
&\hyperref[acr:icpms]{ICP-MS}  & \hyperref[acr:pnnl]{PNNL}
& 4.47+-0.27              &                           &  2.3+-0.4                &                           \\
\sample{R-120.1.1}\phantom{$^\dagger$}   & \hyperref[acr:qep]{QEP} (MCAM) TIVAR 1000 \hyperref[acr:esd]{ESD}, C-filled \hyperref[acr:uhmw]{UHMW} \hyperref[acr:pe]{PE}%
&\hyperref[acr:icpms]{ICP-MS}  & \hyperref[acr:pnnl]{PNNL}
& 224+-32                 &                           &  10.1+-1.4               &                           \\
\sample{R-201.1.1}\phantom{$^\dagger$}   & \hyperref[acr:mcam]{MCAM} TIVAR \hyperref[acr:esd]{ESD}, C-filled \hyperref[acr:uhmw]{UHMW} \hyperref[acr:pe]{PE} (SUHMWAS1.250BK)%
&\hyperref[acr:agge]{A.G.\,Ge} & \hyperref[acr:ua]{UA}
&                          &  300+-130                &  730+-260                &  820+-270                \\
\sample{R-200.1.1}\phantom{$^\dagger$}   & \hyperref[acr:mcam]{MCAM} TIVAR \hyperref[acr:ec]{EC} (CS1430-02000)%
&\hyperref[acr:agge]{A.G.\,Ge} & \hyperref[acr:ua]{UA}
&                          &  <420                     &  <2000                    &  3800+-700               \\
\hline%
& \textbf{Structural polymer}\\
\sample{R-143.1.1}\phantom{$^\dagger$}   & Solvay Torlon 4200 \hyperref[acr:pai]{PAI} (\hyperref[acr:dp]{DP} TR 4200 9\mm)%
&\hyperref[acr:naa]{NAA}       & \hyperref[acr:ua]{UA}
& <190                     &                           &  <180                     &  107+-28                 \\
\sample{R-144.1.1}\phantom{$^\dagger$}   & Solvay Torlon 4203 \hyperref[acr:pai]{PAI} (\hyperref[acr:dp]{DP} TR 4203 4\mm)%
&\hyperref[acr:naa]{NAA}       & \hyperref[acr:ua]{UA}
& <550                     &                           &  <450                     &  260+-50                 \\
\sample{R-146.1.1}\phantom{$^\dagger$}   & \hyperref[acr:apt]{APT} Natural Semilon \hyperref[acr:pei]{PEI}, Sabic Ultem 1000 (\hyperref[acr:pp]{PP} RULT1000NA.250)%
&\hyperref[acr:naa]{NAA}       & \hyperref[acr:ua]{UA}
& 21+-13                  &                           &  84+-10                  &  48+-7                   \\
\sample{R-147.1.1}\phantom{$^\dagger$}   & Ensinger TECAPEI \hyperref[acr:pei]{PEI}, Sabic Ultem 1000 (\hyperref[acr:ip]{IP} RR00250x120000)%
&\hyperref[acr:naa]{NAA}       & \hyperref[acr:ua]{UA}
& <15                      &                           &  9.5+-2.3                &  15.3+-3.0               \\
\sample{R-147.1.2}\phantom{$^\dagger$}   & Ensinger TECAPEI \hyperref[acr:pei]{PEI}, Sabic Ultem 1000 (\hyperref[acr:ip]{IP} RR00250x120000)%
&\hyperref[acr:naa]{NAA}       & \hyperref[acr:ua]{UA}
& <44                      &                           &  28.6+-3.4               &  53+-8                   \\
\sample{R-145.1.1}\phantom{$^\dagger$}   & \hyperref[acr:mcam]{MCAM} Duratron T4203 \hyperref[acr:pai]{PAI} (\hyperref[acr:pp]{PP} RTOR4203.250)%
&\hyperref[acr:naa]{NAA}       & \hyperref[acr:ua]{UA}
& <580                     &                           &  <28                      &  290+-140                \\
\sample{R-184.1.1}\phantom{$^\dagger$}   & \hyperref[acr:mcam]{MCAM} Duratron U1000 \hyperref[acr:pei]{PEI}, Sabic Ultem 1000 (\hyperref[acr:bp]{BP} aR478-00187)%
&\hyperref[acr:naa]{NAA}       & \hyperref[acr:ua]{UA}
& <18                      &                           &  67+-7                   &  106+-12                 \\
\hline%
& \textbf{Solder} \\
\sample{R-152.1.1}\phantom{$^\dagger$}   & Nippon Micrometal Corporation AuSn 100\micron\ spheres (LF60)%
&\hyperref[acr:icpms]{ICP-MS}  & \hyperref[acr:pnnl]{PNNL}
& 90+-20                  &                           &  68+-14                  &                           \\
\cline{2-2}%
&\qquad Sn96.5/Ag3.0/Cu0.5 \\
\sample{R-127.3.1}$^\dagger$             & Kester lead-free, no flux (14-7068-0062)%
&\hyperref[acr:ugge]{U.G.~Ge}  & \hyperref[acr:surf]{SURF}
&                          &  280+-60                 &  <75                      &  230+-100                \\
\sample{R-127.4.1}$^\dagger$             & Kester lead-free, no flux (14-7068-0062)%
&\hyperref[acr:agge]{A.G.\,Ge} & \hyperref[acr:ua]{UA}
&                          &  1100+-500               &  <410                     &  <1300                    \\
\sample{R-127.5.1}\phantom{$^\dagger$}   & Kester lead-free, no flux (14-7068-0062)%
&\hyperref[acr:icpms]{ICP-MS}  & \hyperref[acr:pnnl]{PNNL}
& 3.3+-0.8                &                           &  2.9+-1.0                &                           \\
\sample{R-128.3.1}$^\dagger$             & Alpha lead-free, no flux (147002)%
&\hyperref[acr:ugge]{U.G.~Ge}  & \hyperref[acr:surf]{SURF}
&                          &  <94                      &  <120                     &  <180                     \\
\sample{R-128.4.1}$^\dagger$             & Alpha lead-free, no flux (147002)%
&\hyperref[acr:agge]{A.G.\,Ge} & \hyperref[acr:ua]{UA}
&                          &  1500+-600               &  <580                     &  <820                     \\
\sample{R-128.5.1}\phantom{$^\dagger$}   & Alpha lead-free, no flux (147002)%
&\hyperref[acr:icpms]{ICP-MS}  & \hyperref[acr:pnnl]{PNNL}
& 4.2+-1.5                &                           &  2.3+-1.3                &                           \\
\sample{R-135.1.1}$^\dagger$             & Alpha lead-free, flux core (24-7068-6422)%
&\hyperref[acr:ugge]{U.G.~Ge}  & \hyperref[acr:surf]{SURF}
&                          &  180+-100                &  <100                     &  1160+-250               \\
\sample{R-135.2.1}$^\dagger$             & Alpha lead-free, flux core (24-7068-6422)%
&\hyperref[acr:agge]{A.G.\,Ge} & \hyperref[acr:ua]{UA}
&                          &  <480                     &  <310                     &  <2100                    \\
        \end{tblr}%
    \end{table*}%
\end{turnpage}%
\begin{turnpage}%
    \begin{table*}[t]%
        \begin{tblr}{width=\linewidth,colsep=2pt,colspec={%
                @{}r|%
                M%
                @{}l%
                l%
                @{}Q[c, si={table-align-comparator=false,table-format=<4.2+-1.2e1}]%
                @{}Q[c, si={table-align-comparator=false,table-format=<4.2+-1.2e1}]%
                @{}Q[c, si={table-align-comparator=false,table-format=<4.2+-1.2e1}]%
                @{}Q[c, si={table-align-comparator=false,table-format=<4.2+-1.2e1}]%
            }%
        }%
\hline%
& \textbf{Wires}\\
\sample{R-035.1.1}$^\dagger$             & W.L.~Gore \hyperref[acr:awg]{AWG} 28 Cu conductor, 0.015\inch\ Teflon insulation (012749900)%
&\hyperref[acr:ugge]{U.G.~Ge}  & \hyperref[acr:vda]{VdA}
&                          &  <44                      &  <51                      &  <150                     \\
\sample{R-126.1.1}\phantom{$^\dagger$}   & Ametek Al bonding, Al-1\%\,Si alloy (109433/A4)%
&\hyperref[acr:icpms]{ICP-MS}  & \hyperref[acr:pnnl]{PNNL}
& 1.4+-0.4e4              &                           &  4.7+-1.3e4              &                           \\
\sample{R-134.1.1}\phantom{$^\dagger$}   & Ametek gold bonding, 99.99\% Au (10190)%
&\hyperref[acr:icpms]{ICP-MS}  & \hyperref[acr:pnnl]{PNNL}
& <230                     &                           &  26+-8                   &                           \\
\sample{R-149.1.1}\phantom{$^\dagger$}   & Hamamatsu Sn-Au bump bonding%
&\hyperref[acr:icpms]{ICP-MS}  & \hyperref[acr:pnnl]{PNNL}
& 22+-9                   &                           &  6.0+-3.0                &                           \\
\sample{R-150.1.1}\phantom{$^\dagger$}   & Hamamatsu Sn-Cu bump bonding%
&\hyperref[acr:icpms]{ICP-MS}  & \hyperref[acr:pnnl]{PNNL}
& 22+-6                   &                           &  <14                      &                           \\
\hline%
& \textbf{Polyimide flex cables} \\
\sample{R-073.1.1}$^\dagger$             & McMaster-Carr Kapton film (2271K5)%
&\hyperref[acr:agge]{A.G.\,Ge} & \hyperref[acr:ua]{UA}
&                          &  <4400                    &  1.8+-0.4e4              &  <3700                    \\
\sample{R-093.1.1}\phantom{$^\dagger$}   & CS Hyde DuPont Kapton HN (18-5F-12\inch) \cite{ARNQUIST2020163573}%
&\hyperref[acr:icpms]{ICP-MS}  & \hyperref[acr:pnnl]{PNNL}
& 921+-11                 &                           &  86.0+-1.3               &                           \\
\sample{R-093.1.2}\phantom{$^\dagger$}   & CS Hyde DuPont Kapton HN (18-5F-12\inch)%
&\hyperref[acr:icpms]{ICP-MS}  & \hyperref[acr:ihep]{IHEP}
& 961+-12                 &                           &  77.0+-2.0               &                           \\
\sample{R-093.1.3}\phantom{$^\dagger$}   & CS Hyde DuPont Kapton HN (18-5F-12\inch)%
&\hyperref[acr:naa]{NAA}       & \hyperref[acr:ua]{UA}
& 1300+-90                &                           &  140+-70                 &                           \\
\sample{R-086.1.1}\phantom{$^\dagger$}   & DuPont Pyralux AP Cu-clad laminate, double-sided (8535R)  \cite{ARNQUIST2020163573}%
&\hyperref[acr:icpms]{ICP-MS}  & \hyperref[acr:pnnl]{PNNL}
& 158+-6                  &                           &  24.1+-0.9               &                           \\
\sample{R-092.3.1}\phantom{$^\dagger$}   & DuPont Kapton 100HN \cite{ARNQUIST2020163573}%
&\hyperref[acr:icpms]{ICP-MS}  & \hyperref[acr:pnnl]{PNNL}
& 966+-14                 &                           &  89.8+-3.2               &                           \\
\sample{R-116.1.1}\phantom{$^\dagger$}   & DuPont Kapton 100HN \cite{ARNQUIST2020163573}%
&\hyperref[acr:icpms]{ICP-MS}  & \hyperref[acr:pnnl]{PNNL}
& 931+-5                  &                           &  115.0+-1.2              &                           \\
\sample{R-092.1.1}\phantom{$^\dagger$}   & DuPont Kapton 300HN \cite{ARNQUIST2020163573}%
&\hyperref[acr:icpms]{ICP-MS}  & \hyperref[acr:pnnl]{PNNL}
& 1080+-40                &                           &  250+-8                  &  44+-18                   \\
\sample{R-092.2.1}\phantom{$^\dagger$}   & DuPont Kapton 500HN \cite{ARNQUIST2020163573}%
&\hyperref[acr:icpms]{ICP-MS}  & \hyperref[acr:pnnl]{PNNL}
& 830+-70                 &                           &  73+-7                   &                           \\
\sample{R-094.1.1}\phantom{$^\dagger$}   & DuPont Kapton 200DR9, C-loaded%
&\hyperref[acr:icpms]{ICP-MS}  & \hyperref[acr:pnnl]{PNNL}
& 58+-10                  &                           &  45.2+-2.3               &                           \\
\sample{R-094.2.1}\phantom{$^\dagger$}   & DuPont Kapton 300DR12, C-loaded%
&\hyperref[acr:icpms]{ICP-MS}  & \hyperref[acr:pnnl]{PNNL}
& 6900+-500               &                           &  4880+-170               &                           \\
\sample{R-098.1.1}\phantom{$^\dagger$}   & DuPont Kapton 100XC, C-loaded%
&\hyperref[acr:icpms]{ICP-MS}  & \hyperref[acr:pnnl]{PNNL}
& 1.62+-0.11e4            &                           &  1.70+-0.04e4            &                           \\
\sample{R-118.1.1}\phantom{$^\dagger$}   & DuPont Kapton 200HH, R\&D line \cite{ARNQUIST2020163573}%
&\hyperref[acr:icpms]{ICP-MS}  & \hyperref[acr:pnnl]{PNNL}
& 12.3+-1.9               &                           &  18.5+-2.3               & 34+-14                          \\
\sample{R-117.1.1}\phantom{$^\dagger$}   & DuPont Kapton XC KSW, C-loaded, R\&D line%
&\hyperref[acr:icpms]{ICP-MS}  & \hyperref[acr:pnnl]{PNNL}
& 4.5+-0.9e4              &                           &  2.2+-0.6e4              &                           \\
\sample{R-087.1.1}\phantom{$^\dagger$}   & Sheldahl Novaclad, Cu clad, double-sided (146319-009) \cite{ARNQUIST2020163573}%
&\hyperref[acr:icpms]{ICP-MS}  & \hyperref[acr:pnnl]{PNNL}
& 283+-21                 &                           &  50+-4                   &                           \\
\sample{R-099.1.1}\phantom{$^\dagger$}   & Sheldahl/Multek 1000A 100CB, Ge-coated (147361)%
&\hyperref[acr:icpms]{ICP-MS}  & \hyperref[acr:pnnl]{PNNL}
& 830+-50                 &                           &  627+-10                 &                           \\
\sample{R-099.2.1}\phantom{$^\dagger$}   & Sheldahl/Multek 1500A HN (160759-009)%
&\hyperref[acr:icpms]{ICP-MS}  & \hyperref[acr:pnnl]{PNNL}
& <1.8e4                   &                           &  115+-7                  &                           \\
\sample{R-095.1.1}\phantom{$^\dagger$}   & Fralock XC (0.001\inch), C-loaded%
&\hyperref[acr:icpms]{ICP-MS}  & \hyperref[acr:pnnl]{PNNL}
& 1.2+-0.4e4              &                           &  1.4+-0.4e4              &                           \\
\sample{R-103.1.1}\phantom{$^\dagger$}   & Fralock Cirlex, Cu-clad (8010121083) \cite{ARNQUIST2020163573}%
&\hyperref[acr:icpms]{ICP-MS}  & \hyperref[acr:pnnl]{PNNL}
& 410+-50                 &                           &  71.4+-2.1               &                           \\
\cline{2-2}%
& \qquad QFlex Taiflex custom two-layer cables (2FPDR2005JA)\\
& \qquad\qquad Batch 1\\
\sample{R-137.1.1}\phantom{$^\dagger$}   & Original cable (uncleaned)%
&\hyperref[acr:icpms]{ICP-MS}  & \hyperref[acr:pnnl]{PNNL}
& 6200+-100               &                           &  63+-5                   &                           \\
\sample{R-137.2.1}\phantom{$^\dagger$}   & Base material (copper clad \hyperref[acr:pi]{PI}) \cite{Arnquist2023}%
&\hyperref[acr:icpms]{ICP-MS}  & \hyperref[acr:pnnl]{PNNL}
& 9+-4                    &                           &  8+-6                    &                           \\
%
%\sample{J-219}\phantom{$^\dagger$}   & Base material (copper clad \hyperref[acr:pi]{PI})%
%&\hyperref[acr:ugge]{U.G.~Ge}  & SNOLAB
%& 17+-4                    &                           &  104+-16                & 216+-20                   \\
%
\sample{R-137.3.1}\phantom{$^\dagger$}   & \hyperref[acr:pnnl]{PNNL} cleaned cable%
&\hyperref[acr:icpms]{ICP-MS}  & \hyperref[acr:pnnl]{PNNL}
& 60+-20                  &                           &  36.0+-3.0               &                           \\
%
%
%
%\cline{2-2}
& \qquad\qquad Batch 2\\%2\mil\ \hyperref[acr:pi]{PI}, 18\micron\ copper on both sides\\
\sample{R-140.2.1}\phantom{$^\dagger$}   & Base material (copper clad \hyperref[acr:pi]{PI})%
&\hyperref[acr:icpms]{ICP-MS}  & \hyperref[acr:pnnl]{PNNL}
& 3.9+- 0.9                &                           &  <9.0                     &                           \\
\sample{R-140.1.1}\phantom{$^\dagger$}   & Finished uncleaned cable%
&\hyperref[acr:icpms]{ICP-MS}  & \hyperref[acr:pnnl]{PNNL}
& 1300+- 300               &                           &  16+- 6                   &                           \\
\sample{R-140.3.1}\phantom{$^\dagger$}   & Finished uncleaned cable, cleaned by QFlex%
&\hyperref[acr:icpms]{ICP-MS}  & \hyperref[acr:pnnl]{PNNL}
& 18.0+- 1.0               &                           &  18+- 5                   &                           \\
\sample{R-140.4.1}\phantom{$^\dagger$}   & Copper from base material (normalized to laminate mass)%
&\hyperref[acr:icpms]{ICP-MS}  & \hyperref[acr:pnnl]{PNNL}
& 4.0+- 0.7                &                           &  1.2+- 0.7                &                           \\
\sample{R-140.5.1}\phantom{$^\dagger$}   & Copper from uncleaned cable (normalized to laminate mass)%
&\hyperref[acr:icpms]{ICP-MS}  & \hyperref[acr:pnnl]{PNNL}
& 1500+- 300               &                           &  17+- 6                   &                           \\
\sample{R-140.6.1}\phantom{$^\dagger$}   & Copper from cleaned cable (normalized to laminate mass)%
&\hyperref[acr:icpms]{ICP-MS}  & \hyperref[acr:pnnl]{PNNL}
& 12.0+- 3.0               &                           &  12+- 4                   &                           \\
\sample{R-140.7.1}\phantom{$^\dagger$}   & \hyperref[acr:pi]{PI} from base material (normalized to laminate mass)%
&\hyperref[acr:icpms]{ICP-MS}  & \hyperref[acr:pnnl]{PNNL}
& 0.90+- 0.20              &                           &  1.1+- 0.5                &                           \\
\sample{R-140.8.1}\phantom{$^\dagger$}   & \hyperref[acr:pi]{PI} from uncleaned cable (normalized to laminate mass)%
&\hyperref[acr:icpms]{ICP-MS}  & \hyperref[acr:pnnl]{PNNL}
& 1.90+- 0.30              &                           &  2.5+- 0.9                &                           \\
\sample{R-140.9.1}\phantom{$^\dagger$}   & \hyperref[acr:pi]{PI} from cleaned cable (normalized to laminate mass)%
&\hyperref[acr:icpms]{ICP-MS}  & \hyperref[acr:pnnl]{PNNL}
& 1.60+- 0.30              &                           &  1.8+- 0.4                &                           \\
\hline%
& \textbf{Other cables} \\
\sample{R-028.1.1}$^\dagger$             & 3M flat cable (SL8802-/08-201N5\_CUT)%
&\hyperref[acr:agge]{A.G.\,Ge} & \hyperref[acr:ua]{UA}
&                          &  530+- 340                &  6900+- 900               & 2500+-400                 \\
\sample{R-029.1.1}$^\dagger$             & Cinch Ethernet cable, Cat 6 (73-8891-25)%
&\hyperref[acr:agge]{A.G.\,Ge} & \hyperref[acr:ua]{UA}
&                          &  <330                     &  1800+- 1100              & <840                      \\
\sample{R-174.1.1}\phantom{$^\dagger$}   & 3M twin axial, cable only (8F36-AAA105-XXX-ND)%
&\hyperref[acr:agge]{A.G.\,Ge} & \hyperref[acr:ua]{UA}
& 1.34+- 0.33e5            &  900+- 400                &  <2300                    & 3.09+-0.34e4              \\
\sample{R-174.2.1}\phantom{$^\dagger$}   & 3M twin axial, connectors (8F36-AAA105-XXX-ND)%
&\hyperref[acr:agge]{A.G.\,Ge} & \hyperref[acr:ua]{UA}
& 6.8+- 2.0e5              &  8.7+- 0.9e5              &  2.43+- 0.24e6            & 3.8+-0.4e5                \\
\sample{R-178.1.1}\phantom{$^\dagger$}   & Meritec twin axial, cable only (948241)%
&\hyperref[acr:agge]{A.G.\,Ge} & \hyperref[acr:ua]{UA}
& <7.2e4                   &  <860                     &  <3100                    & 2200+-1000                \\
\sample{R-178.2.1}\phantom{$^\dagger$}   & Meritec twin axial, connectors (948241)%
&\hyperref[acr:agge]{A.G.\,Ge} & \hyperref[acr:ua]{UA}
& <2.7e5                   &  2.99+- 0.31e5            &  1.55+- 0.16e6            & 1.65+-0.21e5              \\
\sample{R-176.1.1}\phantom{$^\dagger$}   & Samtex twin axial, cable only (HDR-218107-04-NVACE)%
&\hyperref[acr:agge]{A.G.\,Ge} & \hyperref[acr:ua]{UA}
& <3.1e4                   &  <250                     &  <560                     & 3200+-1100                \\
\sample{R-176.2.1}\phantom{$^\dagger$}   & Samtex twin axial, connectors (HDR-218107-04-NVACE)%
&\hyperref[acr:agge]{A.G.\,Ge} & \hyperref[acr:ua]{UA}
& <1.6e5                   &  3.4+- 0.4e4              &  1.89+- 0.20e5            & 2.5+-0.5e4                \\
\sample{R-173.1.1}\phantom{$^\dagger$}   & Samtec custom twin axial, cable only (HDR-192192-03)%
&\hyperref[acr:agge]{A.G.\,Ge} & \hyperref[acr:ua]{UA}
& <3.6e4                   &  <230                     &  <1800                    & 3900+-900                 \\
\sample{R-173.2.1}\phantom{$^\dagger$}   & Samtec custom twin axial, connectors (HDR-192192-03)%
&\hyperref[acr:agge]{A.G.\,Ge} & \hyperref[acr:ua]{UA}
& <1.7e6                   &  9.8+- 1.0e5              &  4.5+- 0.4e6              & 4.8+-0.6e5                \\
        \end{tblr}%
    \end{table*}%
\end{turnpage}%
\begin{turnpage}%
    \begin{table*}[t]
        \begin{tblr}{width=\linewidth,colsep=2pt,colspec={%
                @{}r|%
                M%
                @{}l%
                l%
                @{}Q[c, si={table-align-comparator=false,table-format=<4.2+-1.2e1}]%
                @{}Q[c, si={table-align-comparator=false,table-format=<4.2+-1.2e1}]%
                @{}Q[c, si={table-align-comparator=false,table-format=<4.2+-1.2e1}]%
                @{}Q[c, si={table-align-comparator=false,table-format=<4.2+-1.2e1}]%
            }%
        }%
\hline%
& \textbf{Photodetector}\\
\sample{R-003.1.1}\phantom{$^\dagger$}   & \hyperref[acr:rmd]{RMD} \hyperref[acr:sipm]{SiPM}%
&\hyperref[acr:naa]{NAA}       & \hyperref[acr:ua]{UA}
& <420                     &                           &  <120                     & <8.7                      \\
\sample{R-003.2.1}\phantom{$^\dagger$}   & \hyperref[acr:fbk]{FBK} \hyperref[acr:sipm]{SiPM}, batch 1%
&\hyperref[acr:naa]{NAA}       & \hyperref[acr:ua]{UA}
& <130                     &                           &  <210                     & <5.4                      \\
\sample{R-003.2.2}\phantom{$^\dagger$}   & \hyperref[acr:fbk]{FBK} \hyperref[acr:sipm]{SiPM}, batch 2%
&\hyperref[acr:naa]{NAA}       & \hyperref[acr:ua]{UA}
& <20                      &                           &  <4.6                     & <16                       \\
\sample{R-003.2.3}\phantom{$^\dagger$}   & \hyperref[acr:fbk]{FBK} \hyperref[acr:sipm]{SiPM}, batch 3%
&\hyperref[acr:naa]{NAA}       & \hyperref[acr:ua]{UA}
& <2.0                     &                           &  3.15+- 0.23              & <4.7                      \\
\sample{R-003.2.4}\phantom{$^\dagger$}   & \hyperref[acr:fbk]{FBK} \hyperref[acr:sipm]{SiPM}, batch 4%
&\hyperref[acr:naa]{NAA}       & \hyperref[acr:ua]{UA}
& <11                      &                           &  0.17+- 0.12              & 3.8+-0.4                  \\

\sample{R-076.1.1}\phantom{$^\dagger$}   & \hyperref[acr:fbk]{FBK} low field \hyperref[acr:sipm]{SiPM} (63XFBK)%
&\hyperref[acr:icpms]{ICP-MS}  & \hyperref[acr:pnnl]{PNNL}
& 0.86+- 0.05              &                           &  0.45+- 0.12              &                           \\
\sample{R-076.1.2}\phantom{$^\dagger$}   & \hyperref[acr:fbk]{FBK} low field \hyperref[acr:sipm]{SiPM} (63XFBK)%
&\hyperref[acr:naa]{NAA}       & \hyperref[acr:ua]{UA}
& <68                      &                           &  <22                      &                           \\
\sample{R-076.2.1}\phantom{$^\dagger$}   & \hyperref[acr:fbk]{FBK} standard field \hyperref[acr:sipm]{SiPM} (64XFBK)%
&\hyperref[acr:icpms]{ICP-MS}  & \hyperref[acr:pnnl]{PNNL}
& 0.986+- 0.014            &                           &  0.44+- 0.05              &                           \\
\sample{R-076.2.2}\phantom{$^\dagger$}   & \hyperref[acr:fbk]{FBK} standard field \hyperref[acr:sipm]{SiPM} (64XFBK)%
&\hyperref[acr:icpms]{ICP-MS}  & \hyperref[acr:ihep]{IHEP}
& 0.070+- 0.020            &                           &  0.110+- 0.030            &                           \\
\cline{2-2}%
& \qquad Silicon wafers \\
\sample{R-024.1.1}$^\dagger$             & BNL pulverized%
&\hyperref[acr:agge]{A.G.\,Ge} & \hyperref[acr:ua]{UA}
&                          &  <300                     &  <230                     & <2400                     \\
\sample{R-024.1.4}\phantom{$^\dagger$}   & BNL pulverized%
&\hyperref[acr:icpms]{ICP-MS}  & \hyperref[acr:pnnl]{PNNL}
& 13.20+- 0.10             &                           &  25.7+- 0.7               &                           \\
\sample{R-031.1.1}\phantom{$^\dagger$}   & \hyperref[acr:fbk]{FBK} P1B blank%
&\hyperref[acr:naa]{NAA}       & \hyperref[acr:ua]{UA}
& <0.29                    &                           &  <0.67                    & 1.82+-0.23                \\
\sample{R-031.1.4}\phantom{$^\dagger$}   & \hyperref[acr:fbk]{FBK} P1B blank%
&\hyperref[acr:icpms]{ICP-MS}  & \hyperref[acr:pnnl]{PNNL}
& <0.62                    &                           &  0.501+- 0.032            &                           \\
\sample{R-031.3.1}\phantom{$^\dagger$}   & \hyperref[acr:fbk]{FBK} P3B-Ti%
&\hyperref[acr:naa]{NAA}       & \hyperref[acr:ua]{UA}
& <0.60                    &                           &  0.35+- 0.15              & 0.60+-0.13                \\
\sample{R-077.2.1}\phantom{$^\dagger$}   & Teledyne DALSA Semiconductor blank%
&\hyperref[acr:icpms]{ICP-MS}  & \hyperref[acr:pnnl]{PNNL}
& 0.62+- 0.11              &                           &  0.84+- 0.06              &                           \\
\sample{R-077.1.1}\phantom{$^\dagger$}   & Teledyne DALSA Semiconductor \hyperref[acr:cmos]{CMOS}%
&\hyperref[acr:icpms]{ICP-MS}  & \hyperref[acr:pnnl]{PNNL}
& 0.20+- 0.09              &                           &  0.11+- 0.04              &                           \\
\sample{R-096.1.1}\phantom{$^\dagger$}   & Hamamatsu VUV3 \hyperref[acr:mppc]{MPPC} without tungsten%
&\hyperref[acr:icpms]{ICP-MS}  & \hyperref[acr:pnnl]{PNNL}
& 0.72+- 0.19              &                           &  0.31+- 0.20              &                           \\
\sample{R-096.1.2}\phantom{$^\dagger$}   & Hamamatsu VUV3 \hyperref[acr:mppc]{MPPC} without tungsten%
&\hyperref[acr:icpms]{ICP-MS}  & \hyperref[acr:ihep]{IHEP}
& 0.210+- 0.030            &                           &  0.310+- 0.030            &                           \\
\sample{R-148.1.1}\phantom{$^\dagger$}   & 21 Hamamatsu, 625\micron\ Si, 0--2\micron\ W, 0--0.2\micron\ TiN (S13375-SI-WAFER)%
&\hyperref[acr:icpms]{ICP-MS}  & \hyperref[acr:pnnl]{PNNL}
& 0.24+-0.04              &                           &  <15                      &                           \\
\sample{R-156.1.1}\phantom{$^\dagger$}   & \hyperref[acr:ime]{IME} custom interposer (2020)%
&\hyperref[acr:icpms]{ICP-MS}  & \hyperref[acr:pnnl]{PNNL}
& 4.6+-0.9                &                           &  2.3+-0.9                &                           \\
\sample{R-158.1.1}\phantom{$^\dagger$}   & Sherbrooke-\hyperref[acr:izm]{IZM} polyimide layer%
&\hyperref[acr:icpms]{ICP-MS}  & \hyperref[acr:pnnl]{PNNL}
& <7.3                     &                           &  12+-7                   &                           \\
\sample{R-158.2.1}\phantom{$^\dagger$}   & Sherbrooke-\hyperref[acr:izm]{IZM} AlSi layer%
&\hyperref[acr:icpms]{ICP-MS}  & \hyperref[acr:pnnl]{PNNL}
& <0.70                    &                           &  1.9+-0.4                &                           \\
\sample{R-158.3.1}\phantom{$^\dagger$}   & Sherbrooke-\hyperref[acr:izm]{IZM} copper layer%
&\hyperref[acr:icpms]{ICP-MS}  & \hyperref[acr:pnnl]{PNNL}
& 0.70+-0.10              &                           &  3.3+-0.8                &                           \\
\sample{R-205.1.1}\phantom{$^\dagger$}   & \hyperref[acr:fbk]{FBK} benzocyclobutene film%
&\hyperref[acr:icpms]{ICP-MS}  & \hyperref[acr:pnnl]{PNNL}
& 55.9+-3.2               &                           &  211+-15                 &  <1300                    \\
\hline%
& \textbf{\hyperref[acr:asic]{ASIC}} \\
\sample{R-019.1.1}$^\dagger$             & MicroBooNE cryogenic%
&\hyperref[acr:ugge]{U.G.~Ge}  & \hyperref[acr:vda]{VdA}
&                          &  1.00+-0.29e4            &  <9800                    &  2.5+-1.3e4              \\
\sample{R-090.1.1}\phantom{$^\dagger$}   & Global Foundry 350\nm\ (NTO-LCSA-GF 0.35 ANALOG-20141120)%
&\hyperref[acr:icpms]{ICP-MS}  & \hyperref[acr:pnnl]{PNNL}
& 0.35+-0.13              &                           &  1.3+-0.7                &                           \\
\sample{R-001.2.1}\phantom{$^\dagger$}   & \hyperref[acr:tsmc]{TSMC} \hyperref[acr:cmos]{CMOS}, standard 0.25\micron\ \hyperref[acr:vcsel]{VCSEL} 1--2\inch%
&\hyperref[acr:icpms]{ICP-MS}  & \hyperref[acr:lu]{LU}
& <850                     &                           &  <59                      &                           \\
\sample{R-001.3.1}\phantom{$^\dagger$}   & \hyperref[acr:tsmc]{TSMC} \hyperref[acr:cmos]{CMOS}, standard 0.25\micron\ PIN diode 3--4\inch%
&\hyperref[acr:icpms]{ICP-MS}  & \hyperref[acr:lu]{LU}
& <1300                    &                           &  <300                     &                           \\
\sample{R-001.8.1}$^\dagger$             & \hyperref[acr:tsmc]{TSMC} \hyperref[acr:cmos]{CMOS}, standard 0.25\micron\  assorted chips%
&\hyperref[acr:ugge]{U.G.~Ge}  & \hyperref[acr:vda]{VdA}
&                          &  <4.4e4                   &  <1.8e5                   &                           \\
\sample{R-089.1.1}\phantom{$^\dagger$}   & \hyperref[acr:tsmc]{TSMC} 180\nm\ LAr cold \hyperref[acr:fe]{FE} \hyperref[acr:asic]{ASIC} (V68T\#2-AA)%
&\hyperref[acr:icpms]{ICP-MS}  & \hyperref[acr:pnnl]{PNNL}
& 2.2+-0.6                &                           &  <7.0                     &                           \\
\sample{R-091.1.1}\phantom{$^\dagger$}   & \hyperref[acr:tsmc]{TSMC} 130\nm\ (run TMGI86 of 06/16/2015 Fab14)%
&\hyperref[acr:icpms]{ICP-MS}  & \hyperref[acr:pnnl]{PNNL}
& 0.67+-0.30              &                           &  0.97+-0.23              &                           \\
\sample{R-155.1.1}\phantom{$^\dagger$}   & \hyperref[acr:mosis]{MOSIS}/\hyperref[acr:tsmc]{TSMC} 65\nm\ (V98G-AA)%
&\hyperref[acr:icpms]{ICP-MS}  & \hyperref[acr:pnnl]{PNNL}
& <13                      &                           &  <48                      &                           \\
\sample{R-155.1.2}\phantom{$^\dagger$}   & \hyperref[acr:mosis]{MOSIS}/\hyperref[acr:tsmc]{TSMC} 65\nm\ (V98G-AA)%
&\hyperref[acr:icpms]{ICP-MS}  & \hyperref[acr:pnnl]{PNNL}
& 2.000+-0.019            &                           &  1.500+-0.013            &                           \\
\hline%
& \textbf{Electronic components} \\
\sample{R-058.1.1}$^\dagger$             & Mini-Systems 1.8\GOhm\ resistors (WA-5-PG-1807-M-N-S62)%
&\hyperref[acr:agge]{A.G.\,Ge} & \hyperref[acr:ua]{UA}
&                          &  <2.2e5                   &  <2.2e6                   &  <9.5e5                   \\
\sample{R-065.1.1}$^\dagger$             & Canberra field-effect transistor%
&\hyperref[acr:agge]{A.G.\,Ge} & \hyperref[acr:ua]{UA}
&                          &  <7.6e4                   &  <9.1e4                   &  1.1+-0.5e5              \\
\sample{R-061.1.1}$^\dagger$             & \hyperref[acr:mpik]{MPIK} Heidelberg field-effect transistor%
&\hyperref[acr:agge]{A.G.\,Ge} & \hyperref[acr:ua]{UA}
&                          &  3.8+-2.4e4              &  <1.3e5                   &  1.3+-0.4e5              \\
\sample{R-060.1.1}\phantom{$^\dagger$}   & \hyperref[acr:mpik]{MPIK} Heidelberg 1\pF\ feedback capacitors%
&\hyperref[acr:agge]{A.G.\,Ge} & \hyperref[acr:ua]{UA}
& <5.3e5                   &  1.86+-0.06e5            &  8.9+-1.0e4              &  3.3+-1.5e4              \\
\sample{R-059.1.1}\phantom{$^\dagger$}   & American Technical Ceramics 1\pF\ feedback capacitors (100B1R0CAW500-X)%
&\hyperref[acr:agge]{A.G.\,Ge} & \hyperref[acr:ua]{UA}
& <1.7e6                   &  1.36+-0.13e5            &  5.9+-0.4e5              &  <3.0e4                   \\
\cline{2-2}%
& \qquad Silicon capacitors \\
\sample{R-033.1.1}$^\dagger$             & Ipidia (Murata) 100\nF\ (935132425610-T3N)%
&\hyperref[acr:ugge]{U.G.~Ge}  & \hyperref[acr:vda]{VdA}
&                          &  <7.1e4                   &  <3.8e4                   &  <8.7e4                   \\
\sample{R-195.1.1}\phantom{$^\dagger$}   & Murata 3.3\microF\ \hyperref[acr:hssc]{HSSC} 1812 (935131429733-T3N)%
&\hyperref[acr:icpms]{ICP-MS}  & \hyperref[acr:pnnl]{PNNL}
& 31.0+-2.0               &                           &  146+-9                  &                           \\
\sample{R-209.1.1}\phantom{$^\dagger$}   & Murata 100\nF\ (935131424610-T3N)%
&\hyperref[acr:agge]{U.G.\,Ge} & \hyperref[acr:ua]{UA}
& <3.3e7                   &  1.3+-0.8e6                &  <4.6e6                   &  2.6+-1.2e6                \\
\sample{R-209.1.2}\phantom{$^\dagger$}   & Murata 100\nF\ (935131424610-T3N)%
&\hyperref[acr:ugge]{U.G.\,Ge} & SNOLAB
& <1.2e5                   &  <1.6e4                   &  <2.5e5                   &                           \\
\sample{R-197.1.2}\phantom{$^\dagger$}   & Kyocera 470\pF%
&\hyperref[acr:agge]{A.G.\,Ge} & \hyperref[acr:ua]{UA}
& <2.9e6                   &  2.4+-0.5e5              &  2.4+-1.4e5              &  2.3+-0.9e5              \\
\sample{R-197.2.1}\phantom{$^\dagger$}   & Koycera 470\pF%
&\hyperref[acr:icpms]{ICP-MS}  & \hyperref[acr:ihep]{IHEP}
& 3.1+-0.5e5              &                           &  8.5+-0.4e4              &                           \\
\sample{R-215.1.1}\phantom{$^\dagger$}   & Empower Semiconductor 220\nF\ (EC1001P)%
&\hyperref[acr:icpms]{ICP-MS}  & \hyperref[acr:pnnl]{PNNL}
& <2.2                     &                           &  <7.0                     &  8.8+-1.3                \\
\sample{R-216.1.1}\phantom{$^\dagger$}   & Empower Semiconductor 670\nF\ (EC1100P)%
&\hyperref[acr:icpms]{ICP-MS}  & \hyperref[acr:pnnl]{PNNL}
& 36.0+-3.0               &                           &  20+-14                  &  4.7+-2.2                \\
        \end{tblr}%
    \end{table*}%
\end{turnpage}%
\newcommand{\mgftwo}{\ensuremath{\text{MgF}_2}}%
\begin{turnpage}%
    \begin{table*}[t]%
        \begin{tblr}{width=\linewidth,colsep=2pt,colspec={%
                @{}r|%
                M%
                @{}l%
                l%
                @{}Q[c, si={table-align-comparator=false,table-format=<4.2+-1.2e1}]%
                @{}Q[c, si={table-align-comparator=false,table-format=<4.2+-1.2e1}]%
                @{}Q[c, si={table-align-comparator=false,table-format=<4.2+-1.2e1}]%
                @{}Q[c, si={table-align-comparator=false,table-format=<4.2+-1.2e1}]%
            }%
        }%
\cline{2-2}%
& \qquad FR-4 \hyperref[acr:pcb]{PCB}s \\
\sample{R-157.1.1}\phantom{$^\dagger$}   & \hyperref[acr:bnl]{BNL} generic assembly%
&\hyperref[acr:agge]{A.G.\,Ge} & \hyperref[acr:ua]{UA}
& 4.5+-1.0e5              &  6.0+-0.6e5              &  1.82+-0.18e6            &  1.68+-0.17e5            \\
\sample{R-170.1.1}\phantom{$^\dagger$}   & PCBWay.com unassembled  (W148965ASS19)%
&\hyperref[acr:agge]{A.G.\,Ge} & \hyperref[acr:ua]{UA}
& 1.5+-0.9e6              &  8.7+-0.9e5              &  4.7+-0.5e6              &  5.8+-0.6e5              \\
\sample{R-171.1.1}\phantom{$^\dagger$}   & PCBWay.com unassembled (W148965ASS28)%
&\hyperref[acr:agge]{A.G.\,Ge} & \hyperref[acr:ua]{UA}
& 2.2+-0.4e6              &  1.66+-0.17e6            &  7.9+-0.8e6              &  1.00+-0.10e6            \\
\sample{R-172.1.1}\phantom{$^\dagger$}   & Palpilot unassembled%
&\hyperref[acr:agge]{A.G.\,Ge} & \hyperref[acr:ua]{UA}
& 1.09+-0.22e6            &  1.16+-0.12e6            &  5.5+-0.6e6              &  6.4+-0.7e5              \\
\sample{R-175.1.1}\phantom{$^\dagger$}   & Palpilot unassembled (C180001067)%
&\hyperref[acr:agge]{A.G.\,Ge} & \hyperref[acr:ua]{UA}
& 1.97+-0.26e6            &  1.93+-0.19e6            &  8.9+-0.9e6              &  1.09+-0.11e6            \\
\hline%
& \textbf{Heat transfer fluid} \\
\sample{R-181.1.1}$^\dagger$             & 3M NOVEC-7000%
&\hyperref[acr:icpms]{ICP-MS}  & \hyperref[acr:pnnl]{PNNL}
& 0.0034+-0.0005          &                           &  <0.0013                  &                           \\
\sample{R-182.1.1}\phantom{$^\dagger$}   & 3M NOVEC-7200%
&\hyperref[acr:icpms]{ICP-MS}  & \hyperref[acr:pnnl]{PNNL}
& <0.0043                  &                           &  <0.0013                  &                           \\
\sample{R-191.1.1}\phantom{$^\dagger$}   & \hyperref[acr:tmc]{TMC} reclaimed 3M NOVEC-7200%
&\hyperref[acr:icpms]{ICP-MS}  & \hyperref[acr:pnnl]{PNNL}
& 0.0019+-0.0010          &                           &  0.0030+-0.0020          &                           \\
\sample{R-202.1.1}\phantom{$^\dagger$}   & JuFu JF7100 (2023060390)%
&\hyperref[acr:icpms]{ICP-MS}  & \hyperref[acr:pnnl]{PNNL}
& <0.0017                  &                           &  <0.0012                  &                           \\
\sample{R-203.1.1}\phantom{$^\dagger$}   & JuFu HFE6512 (202305158120)%
&\hyperref[acr:icpms]{ICP-MS}  & \hyperref[acr:pnnl]{PNNL}
& 0.0074+-0.0013          &                           &  0.0045+-0.0004          &                           \\
\hline%
& \textbf{Getter materials} \\
\sample{R-169.1.1}\phantom{$^\dagger$}   & %\hyperref[acr:saes]{SAES}
St 707 Pill/4-2/50 getter material from %EXO-200
\hyperref[acr:saes]{SAES} PS4-MT3, opened 2015%
&\hyperref[acr:icpms]{ICP-MS}  & \hyperref[acr:pnnl]{PNNL}
& 1.60+-0.09e5            &                           &  8800+-1300              &                           \\
\sample{R-159.1.1}\phantom{$^\dagger$}   & %\hyperref[acr:saes]{SAES}
St 707 Pill/4-2/50 getter material from %EXO-200
\hyperref[acr:saes]{SAES} PS4-MT3, opened 2021%
&\hyperref[acr:agge]{A.G.\,Ge} & \hyperref[acr:ua]{UA}
& 1.15+-0.20e5            &  <1500                    &  1.88+-0.21e4            &                           \\
\sample{R-159.1.3}$^\dagger$   & %\hyperref[acr:saes]{SAES}
St 707 Pill/4-2/50 getter material from %EXO-200
\hyperref[acr:saes]{SAES} PS4-MT3, opened 2021%
&\hyperref[acr:ugge]{U.G.~Ge}  & SNOLAB
& 10.1+-1.0e4                &  760+-120                &   1.51+-0.09e4            & <340                    \\
\sample{R-159.1.2}\phantom{$^\dagger$}   & %\hyperref[acr:saes]{SAES}
St 707 Pill/4-2/50 getter material from %EXO-200
\hyperref[acr:saes]{SAES} PS4-MT3, opened 2021%
&\hyperref[acr:icpms]{ICP-MS}  & \hyperref[acr:pnnl]{PNNL}
& 7.5+-0.4e4              &                           &  1.00+-0.30e4            &                           \\
\sample{R-164.1.1}\phantom{$^\dagger$}   & ALB Materials ALB-Z001 (99.95\% Zi, \SI{<300}{ppm} Hf) pellets,%
&\hyperref[acr:agge]{A.G.\,Ge} & \hyperref[acr:ua]{UA}
&                          &  <680                     &  <470                     &  <460                     \\
\sample{R-164.1.2}\phantom{$^\dagger$}   & ALB Materials ALB-Z001 (99.95\% Zi, \SI{<300}{ppm} Hf) pellets,%
&\hyperref[acr:icpms]{ICP-MS}  & \hyperref[acr:pnnl]{PNNL}
& 2000+-400               &                           &  300+-200                &                           \\
\sample{R-164.2.1}\phantom{$^\dagger$}   & ALB Materials ALB-Z001 (99.95\% Zi, \SI{<300}{ppm} Hf) pellets,%
&\hyperref[acr:ugge]{U.G.~Ge}  & \hyperref[acr:lngs]{LNGS}
& <1600                    &  <110                     &  540+-150                &  <310                     \\
\sample{R-177.1.1}\phantom{$^\dagger$}   & ALB Materials ALB-Z001 (99.95\% Zi, \SI{<300}{ppm} Hf) \hyperref[acr:gdms]{GD-MS} rod%
&\hyperref[acr:gdms]{GD-MS}    & \hyperref[acr:nrc]{NRC}
& <1000                    &                           &  <800                     &                           \\
\hline%
& \textbf{Miscellaneous} \\
\sample{R-009.1.1}$^\dagger$             & McMaster-Carr \hyperref[acr:uhmw]{UHMW} \hyperref[acr:pe]{PE} for gasket (8701K67)%
&\hyperref[acr:ugge]{U.G.~Ge}  & \hyperref[acr:vda]{VdA}
&                          &  150+-40                 &  <150                     &  <190                     \\
\sample{R-064.1.1}$^\dagger$             & Canberra Mylar sheet%
&\hyperref[acr:agge]{A.G.\,Ge} & \hyperref[acr:ua]{UA}
&                          &  <3500                    &  <1.1e4                   &  6800+-3500              \\
\sample{R-069.1.1}$^\dagger$             & Canberra Teflon%
&\hyperref[acr:agge]{A.G.\,Ge} & \hyperref[acr:ua]{UA}
&                          &  <1300                    &  <8900                    &  <1.4e4                   \\
\sample{R-063.1.1}$^\dagger$             & Canberra tin%
&\hyperref[acr:agge]{A.G.\,Ge} & \hyperref[acr:ua]{UA}
&                          &  <4900                    &  3.5+-1.4e4              &  <1.3e4                   \\
\sample{R-187.1.1}\phantom{$^\dagger$}   & Gold washer for thermal coupling%
&\hyperref[acr:agge]{A.G.\,Ge} & \hyperref[acr:ua]{UA}
& <5.0e5                   &  <2.0e4                   &  7+-5e4                  &  <1.1e4                   \\
\sample{R-066.1.1}$^\dagger$             & Canberra screws (S3) and washers (W1 and W2)%
&\hyperref[acr:agge]{A.G.\,Ge} & \hyperref[acr:ua]{UA}
&                          &  <23                      &  69+-33                  &  <30                      \\
\sample{R-067.1.1}$^\dagger$             & Canberra screws (S3) and washers (W3)%
&\hyperref[acr:agge]{A.G.\,Ge} & \hyperref[acr:ua]{UA}
&                          &  <3000                    &  <8400                    &  <1.6e4                   \\
\sample{R-068.1.1}$^\dagger$             & Canberra screws (S2)%
&\hyperref[acr:agge]{A.G.\,Ge} & \hyperref[acr:ua]{UA}
&                          &  5000+-4000              &  2.3+-1.0e4              &  1.0+-0.6e4              \\
\sample{R-121.1.1}\phantom{$^\dagger$}   & CoorsTek silicon carbide%
&\hyperref[acr:naa]{NAA}       & \hyperref[acr:ua]{UA}
& <0.76                    &                           &  0.58+-0.15              &  0.11+-0.06              \\
\sample{R-122.1.1}\phantom{$^\dagger$}   & Corning high vacuum grease%
&\hyperref[acr:agge]{A.G.\,Ge} & \hyperref[acr:ua]{UA}
&                          &  5.2+-1.1e4              &  1.0+-0.6e4              &  5.6+-2.1e4              \\
\sample{R-125.1.1}\phantom{$^\dagger$}   & DuPont Teijin Melinex 1311 antistatic film, \SI{440}{ga}%
&\hyperref[acr:icpms]{ICP-MS}  & \hyperref[acr:pnnl]{PNNL}
& 9.9+-1.9                &                           &  17.3+-2.0               &                           \\
\sample{R-151.1.1}$^\dagger$             & Merck KGaA Patinal \mgftwo\ granules (\hyperref[acr:casrn]{CASRN} No: 7783-40-6)%
&\hyperref[acr:agge]{A.G.\,Ge} & \hyperref[acr:ua]{UA}
&                          &  <1500                    &  <2400                    &  <1900                    \\
\sample{R-154.1.1}\phantom{$^\dagger$}   & Merck KGaA Patinal \mgftwo\ granules (\hyperref[acr:casrn]{CASRN} No: 7783-40-6)%
&\hyperref[acr:ugge]{U.G.~Ge}  & SNOLAB
&                          &  48+-33                  &  440+-120                &  <110                     \\
\sample{R-180.1.1}\phantom{$^\dagger$}   & SNOLAB dust, J-Drift%
&\hyperref[acr:agge]{A.G.\,Ge} & \hyperref[acr:ua]{UA}
& <1.5e6                   &  3.8+-0.4e5              &  8.3+-1.0e5              &  3.9+-0.4e6              \\
\sample{R-180.3.1}\phantom{$^\dagger$}   & SNOLAB dust, J-Drift (large debris removed)%
&\hyperref[acr:agge]{A.G.\,Ge} & \hyperref[acr:ua]{UA}
& <2.9e6                   &  3.5+-0.8e5              &  4.5+-0.9e5              &  2.48+-0.26e6            \\
\sample{R-180.2.1}\phantom{$^\dagger$}   & SNOLAB dust, Cryopit%
&\hyperref[acr:agge]{A.G.\,Ge} & \hyperref[acr:ua]{UA}
& <2.2e5                   &  5.6+-0.6e4              &  5.0+-0.7e4              &  1.89+-0.20e5            \\
\sample{R-185.1.1}\phantom{$^\dagger$}   & Canberra zeolite molecular sieve type 5A 1/16 pellet%
&\hyperref[acr:agge]{A.G.\,Ge} & \hyperref[acr:ua]{UA}
& 1.61+-0.25e6            &  2.02+-0.20e6            &  7.5+-0.8e6              &  2.34+-0.25e5            \\
\sample{R-186.1.1}\phantom{$^\dagger$}   & Canberra sieve carbon pellet, activated acid washed%
&\hyperref[acr:agge]{A.G.\,Ge} & \hyperref[acr:ua]{UA}
& <7.3e4                   &  3.13+-0.34e4            &  2.21+-0.22e5            &  3.8+-0.4e5              \\
\sample{R-198.1.1}$^\dagger$             & GE GE500 Advanced silicone sealant, clear%
&\hyperref[acr:agge]{A.G.\,Ge} & \hyperref[acr:ua]{UA}
& <3.0e4                   &  <400                     &  <4300                    &  6100+-1700              \\
\sample{R-208.1.2}\phantom{$^\dagger$}   & Phosphor bronze springs%
&\hyperref[acr:icpms]{ICP-MS}  & \hyperref[acr:pnnl]{PNNL}
& 3.6+-2.4                &                           &  5.4+-1.5                &  7.57+-0.32              \\
\sample{R-208.1.1}\phantom{$^\dagger$}   & Phosphor bronze springs bulk (acid etched)%
&\hyperref[acr:icpms]{ICP-MS}  & \hyperref[acr:pnnl]{PNNL}
& <0.35                   &                           &  <2.3                    &  0.21+-0.05              \\
\sample{R-214.1.1.1}\phantom{$^\dagger$} & LOCTITE EA0151 epoxy resin%
&\hyperref[acr:icpms]{ICP-MS}  & \hyperref[acr:pnnl]{PNNL}
& <4.0                     &                           &  <14                      &  <40                      \\
\sample{R-214.1.1.2}\phantom{$^\dagger$} & LOCTITE EA0151 epoxy hardiner%
&\hyperref[acr:icpms]{ICP-MS}  & \hyperref[acr:pnnl]{PNNL}
& 4.3+-0.6                &                           &  6.0+-2.0                &  80+-20                  \\
\sample{R-196.1.1.1}\phantom{$^\dagger$} & Materbond EP29LPSP epoxy resin%
&\hyperref[acr:icpms]{ICP-MS}  & \hyperref[acr:pnnl]{PNNL}
& <5.0                     &                           &  <14                      &  <40                      \\
\sample{R-196.1.1.2}\phantom{$^\dagger$} & Materbond EP29LPSP epoxy hardiner%
&\hyperref[acr:icpms]{ICP-MS}  & \hyperref[acr:pnnl]{PNNL}
& <0.32                    &                           &  <0.70                    &  2.8+-1.9                \\
\hline\hline%
        \end{tblr}%
    \end{table*}%
\end{turnpage}%

%% file: bigtablex.tex
\begingroup%
\squeezetable%
\begin{longtable*}{r|S[table-format=<1.2+-1.2]@{}S[table-space-text-post={},table-align-text-post=false]SS[table-align-text-post=false]@{}S@{}S[table-format=<1.3+-1.3e1]l}%
    \caption{\label{tab:bigtablex}Table of radioassay - additional isotopic data in \textbf{\si{\milli\becquerel/\kilogram}}. Entry \# references the entry \# in Tab.~\ref{tab:bigtable}. The \potwoten\ results are all from $\upalpha$ counting, except for \#126, which is from \gammacounting. $\upalpha$-counting samples are of the same material as those listed in Tab.~\ref{tab:bigtable} but are not the same sample. As detailed in the text, limits marked with \ff\ are more than $3\sigma$ negative.}\\%
    \# & \altwentysix & \cosixty & \agonetenm & \csonethirtyseven & \pbtwoten & \potwoten \\
    \hline\hline%
\ref{R-045.1.1}  &            & 0.102+-0.035 &        &              &              &              \\
\ref{R-045.1.3}  &            & 0.05+-0.04   &        &              &              &              \\
\ref{R-002.7.1}  &            & <0.0054      &        & 0.008+-0.005 &              &              \\
\ref{R-002.7.2}  &            & <0.0066      &        & <0.011       & 710+-70      &              \\
\ref{R-002.8.1}  &            & <0.0033      &        & <0.010       &              &              \\
\ref{R-141.1.1}  & 0.34+-0.18 & 0.70+-0.23   & <0.23  & <0.18        & <5.9         &              \\
\ref{R-075.1.1}  & <0.25      & 4.0+-1.8     &        & 3.0+-1.5     &              &              \\
\ref{R-056.1.1}  & <1.0       & <5.4         &        & <5.6         &              &              \\
\ref{R-056.2.1}  & <0.30      & <3.4         &        & 2.9+-1.5     &              &              \\
\ref{R-056.3.1}  & <2.4       & <3.2         &        & <5.5         &              &              \\
\ref{R-016.1.1}  &            & <4.1         &        & <10          &              &              \\
\ref{R-016.1.2}  &            & <1.3\ff      &        & <2.2         &              &              \\
\ref{R-016.1.3}  &            & <0.11        &        & <0.44        &              &              \\
\ref{R-016.2.1}  &            & <13          &        & <7.6         &              &              \\
\ref{R-016.2.2}  &            & <0.25        &        & <2.1         &              &              \\
\ref{R-016.2.3}  &            & <0.074       &        & <0.27        &              &              \\
\ref{R-016.3.1}  &            & <17          &        & <19          &              &              \\
\ref{R-016.3.2}  &            & <0.14        &        & <1.6         &              &              \\
\ref{R-016.3.3}  &            & <0.078       &        & <0.14        &              &              \\
\ref{R-016.3.4}  &            & <0.0060      &        & <0.094       &              &              \\
\ref{R-017.1.1}  &            & <8.0         &        & <11          &              &              \\
\ref{R-017.1.2}  &            & <0.48        &        & <0.51        &              &              \\
\ref{R-017.1.3}  &            & <0.22        &        & <1.0         &              &              \\
\ref{R-017.2.1}  &            & <2.8         &        & <1.5         &              &              \\
\ref{R-017.2.2}  &            & <0.18\ff     &        & <0.26        &              &              \\
\ref{R-017.2.3}  &            & <0.13        &        & <0.54        &              &              \\
\ref{R-017.3.1}  &            & <0.31        &        & <1.3         &              &              \\
\ref{R-017.3.2}  &            & <0.062       &        & <0.47        &              &              \\
\ref{R-017.3.3}  &            & <0.10        &        & 0.62+-0.23   &              &              \\
\ref{R-039.1.1}  &            & <0.19        &        & <1.8         &              &              \\
\ref{R-041.1.1}  &            & <140         &        & <78          &              &              \\
\ref{R-038.1.1}  &            & <0.41        &        & <2.3         &              &              \\
\ref{R-037.1.1}  &            & <5.5         &        & <10          &              &              \\
\ref{R-036.1.1}  &            & <0.88        &        & <5.0         &              &              \\
\ref{R-054.2.1}  &            & <0.20\ff     &        & <0.099       &              &              \\
\ref{R-054.1.1}  &            & <0.93        &        & <0.67        &              &              \\
\ref{R-062.1.1}  &            & <22          &        & <20          &              &              \\
\ref{R-074.2.1}  &            & <13          &        & <7.0         &              &              \\
\ref{R-074.1.1}  &            & <30\ff       &        & <54          &              &              \\
\ref{R-030.1.1}  &            & <0.77\ff     &        &              &              &              \\
\ref{R-030.2.1}  &            & <0.31        &        & <0.69        &              &              \\
\ref{R-021.1.1}  &            & <1.4         &        & <2.1         &              &              \\
\ref{R-021.1.2}  &            & <0.10        &        & <1.3         &              &              \\
\ref{R-021.2.1}  &            & <0.14        &        & <0.64        &              &              \\
\ref{R-021.3.1}  &            & <0.20        &        & <0.77        &              &              \\
\ref{R-021.8.1}  &            & <0.078       &        & <0.82\ff     &              &              \\
\ref{R-083.1.1}  &            & <0.11        &        & <0.95        &              &              \\
\ref{R-206.2.1}  &            &              &        & 3.0+-1.7     & <140         &              \\
\ref{R-213.3.1}  &            &              &        & 5.2+-1.8     & <0.20        &              \\
\ref{R-127.3.1}  &            & <0.27        & <0.096 & <0.41        & 1.13+-0.11e6 & 9.1+-1.0e5   \\
\ref{R-127.4.1}  &            & <3.9         & <1.8   & <8.1         & 9000+-500    &              \\
\ref{R-128.3.1}  &            & <0.78        & <0.053 & <0.089       &              &              \\
\ref{R-128.4.1}  &            & <3.7         & <1.4   & <9.9         &              &              \\
\ref{R-135.1.1}  &            & <0.20        &        & <0.45        &              &              \\
\ref{R-135.2.1}  &            & 3.6+-2.6     & <2.4   & <12          & <0.20        &              \\
\ref{R-035.1.1}  &            & <0.063       &        & <0.23        &              &              \\
\ref{R-073.1.1}  &            & <14          &        & <26          &              &              \\
\ref{R-028.1.1}  &            & <3.5         &        & <1.6         &              &              \\
\ref{R-029.1.1}  &            & <6.8\ff      &        & <4.0         &              &              \\
\ref{R-024.1.1}  &            & <1.2         &        & <1.9         &              &              \\
\ref{R-019.1.1}  &            & <12          &        & <23          &              &              \\
\ref{R-001.8.1}  &            & <590         &        & <530         &              &              \\
\ref{R-058.1.1}  &            & 4600+-2800   &        & 3300+-1900   &              &              \\
\ref{R-065.1.1}  &            & <360         &        & 370+-280     &              &              \\
\ref{R-061.1.1}  &            & <130         &        & <160         &              &              \\
\ref{R-033.1.1}  &            & <100         &        & <550         &              &              \\
\ref{R-209.1.2}  &            & <680         &        & <5200        &              &              \\
\ref{R-181.1.1}  &            &              &        &              &              & 0.094+-0.023 \\
\ref{R-182.1.1}  &            &              &        &              &              & 0.053+-0.016 \\
\ref{R-191.1.1}  &            &              &        &              &              & <0.018       \\
\ref{R-159.1.3}  &            & <0.88        &        & <0.56 	\\
\ref{R-009.1.1}  &            & <0.32        &        & <0.52        &              &              \\
\ref{R-064.1.1}  &            & <19          &        & <15          &              &              \\
\ref{R-069.1.1}  &            & <5.9         &        & <23          &              &              \\
\ref{R-063.1.1}  &            & <110         &        & 70+-40       &              &              \\
\ref{R-066.1.1}  &            & <0.23        &        & <0.059       &              &              \\
\ref{R-067.1.1}  &            & <23          &        & 25+-11       &              &              \\
\ref{R-068.1.1}  &            & <64          &        & <47          &              &              \\
\ref{R-151.1.1}  &            & 5.8+-3.2     &        & <6.5         &              &              \\
\ref{R-180.1.1}  &            &              &        &              & <5.6e4       &              \\
\ref{R-180.3.1}  &            &              &        &              & <1.1e5       &              \\
\ref{R-185.1.1}  &            &              &        &              & 2.3+-0.6e4   &              \\
\ref{R-198.1.1}  &            &              &        & 30+-9        & <1600        &              \\
\hline\hline%
\end{longtable*}%

%% file: bigtable_radon.tex
\begin{table*}%
    \caption{\label{tab:bigtableradon}Table of $^{222}\rm{Rn}$ emanation measurements in \uBq\ representing the rate of $^{226}\rm{Ra}$ decay supporting the emanation rate from the sample.}%
    \begin{ruledtabular}%
        \begin{tabular*}{\textwidth}{%
            r|r|%
            l%
            l%
            S[table-format=3.3{\sqrm},table-align-text-after=false]%
            S[table-format=<4.1+-3e1,table-align-comparator=false]%
        }%
            R\# & \# & Material & Lab & {Quantity} & {\rntwotwentytwo\ (\si{\micro\becquerel})}\\
            \hline
R1 &                 & BeCu compression spring (10 of Century Spring 10693CS)%
& \hyperref[acr:slac]{SLAC} & 360\cmcm   & 42+-29       \\
\hline && Getter materials\\
R2  & \ref{R-159.1.3} & ST707 Pill/4-2/50 getter material extracted from EXO-200 \hyperref[acr:saes]{SAES} PS4-MT3, opened 2021%
& \hyperref[acr:lu]{LU}     & 410\gram   & 334+-162     \\
R3  & \ref{R-164.2.1} & ALB Materials ALB-Z001 (99.95\% Zr, \SI{<300}{ppm} Hf) pellets, 2x2\mm%
& \hyperref[acr:lu]{LU}     & 501\gram   & 477+-137     \\
R4  & \ref{R-164.2.1} & ALB Materials ALB-Z001 (99.95\% Zr, \SI{<300}{ppm} Hf) pellets, 2x2\mm%
& \hyperref[acr:slac]{SLAC} & 438\gram   & 168+-45      \\
R5  &                 & \hyperref[acr:saes]{SAES} getter PS4-MT50-R-535 (offline/cold)%
& \hyperref[acr:slac]{SLAC} & 1\units    & 176+-48      \\
R6  &                 & \hyperref[acr:saes]{SAES} getter PS4-MT50-R-535 (purifying/hot)%
& \hyperref[acr:slac]{SLAC} & 1\units    & 428+-61      \\
R7  &                 & EXO-200 \hyperref[acr:saes]{SAES} PS4-MT3 purifier (offline/cold)%
& \hyperref[acr:slac]{SLAC} & 1\units    & <73          \\
R8  &                 & EXO-200 \hyperref[acr:saes]{SAES} PS4-MT3 purifier (purifying/hot)%
& \hyperref[acr:slac]{SLAC} & 1\units    & <70          \\
R9  &                 & GetterMax133 copper catalysts pellets%
& \hyperref[acr:slac]{SLAC} & 357\gram   & 1840+-150    \\
\hline && Welds\\
R10  &                 & \hyperref[acr:mig]{MIG} weld coupon, passivated with pickling paste%
& \hyperref[acr:lu]{LU}     & 0.1\sqrm   & <290         \\
R11 &                 & \hyperref[acr:mig]{MIG} weld coupon, unpassivated%
& \hyperref[acr:lu]{LU}     & 0.1\sqrm   & <120         \\
R12 &                 & \hyperref[acr:tig]{TIG} weld coupon, passivated with pickling paste%
& \hyperref[acr:lu]{LU}     & 0.1\sqrm   & <390         \\
R13 &                 & \hyperref[acr:tig]{TIG} weld coupon, unpassivated%
& \hyperref[acr:lu]{LU}     & 0.1\sqrm   & <270         \\
\hline && \hyperref[acr:hv]{HV} cable  \\
R14 & \ref{R-206.2.1} & Dielectric Sciences semi-conductive-\hyperref[acr:pe]{PE} 2353, layers A--C (nEXO)%
& \hyperref[acr:slac]{SLAC} & 10\meter   & <81          \\
R15 & \ref{R-206.2.1} & Dielectric Sciences semi-conductive-\hyperref[acr:pe]{PE} 2353, layers A--C (nEXO)%
& \hyperref[acr:ua]{UA}     & 10.4\meter & <1200        \\
R16 &                 & Dielectric Sciences semi-conductive-\hyperref[acr:pe]{PE} 2353, all layers? (\hyperref[acr:lz]{LZ?})%
& \hyperref[acr:ua]{UA}     & 7.5\meter  & <1800        \\
R17 & \ref{R-213.3.1} & Dielectric Sciences semi-conductive-\hyperref[acr:pe]{PE} 2353, all layers (\hyperref[acr:dune]{DUNE}-\hyperref[acr:nc]{NC})%
& \hyperref[acr:ua]{UA}     & 10.8\meter & <1200        \\
\hline && \multicolumn{4}{l}{\textbf{EXO-200}}\\
R18 &                 & DuPont Teflon TE-6472 fuseable granular resin, lot 0503830033%
& \hyperref[acr:lu]{LU}     & 602.1\gram & 417+-81      \\
R19 &                 & DuPont Teflon TE-6472 fuseable granular resin, lot 0506830001%
& \hyperref[acr:lu]{LU}     & 718.3\gram & <35          \\
R20 &                 & Espanex polyimide laminate cables MC18-25-00CEM%
& \hyperref[acr:lu]{LU}     & 0.8\sqrm   & <116         \\
R21 &                 & CeramTec ceramic electrical breaks (12 of 9998-06-W and 2 of 17199-01-W)%
& \hyperref[acr:lu]{LU}     & {net}      & <95          \\
R22 &                 & GXe valve seats for APTech AP3000 (12 \hyperref[acr:pctfe]{PCTFE} and 12 polyimide)%
& \hyperref[acr:lu]{LU}     & {net}      & <236         \\
R23 &                 & Teflon coated o-rings for Fluitron compressors (4 total)%
& \hyperref[acr:lu]{LU}     & 23.3\cmcm  & <58          \\
R24 &                 & Photo etched phosphor bronze “spider” springs (10 units)%
& \hyperref[acr:lu]{LU}     & 169\cmcm   & <1050        \\
R25 &                 & Photo etched phosphor bronze anode wires plate \#1 scraps%
& \hyperref[acr:lu]{LU}     & 70\cmcm    & <230         \\
R26 &                 & Photo etched phosphor bronze anode wires plate \#2 scraps%
& \hyperref[acr:lu]{LU}     & 82\cmcm    & <490         \\
R27 &                 & Master Bond epoxy EP29LPSP%
& \hyperref[acr:lu]{LU}     & 0.276\sqrm & <140         \\
R28 &                 & Mott filter GSP515H3FF11 (6812039)%
& \hyperref[acr:lu]{LU}     & 1\units    & 570+-70      \\
R29 &                 & McMaster-Carr \hyperref[acr:uhmw]{UHMW} polyethylene 8701K54%
& \hyperref[acr:lu]{LU}     & 0.16\sqrm  & 170+-110     \\
R30 &                 & \hyperref[acr:saes]{SAES} purifier PS4-MT3 \#1%
& \hyperref[acr:lu]{LU}     & 1\units    & <200         \\
R31 &                 & \hyperref[acr:saes]{SAES} purifier PS4-MT3 \#2%
& \hyperref[acr:lu]{LU}     & 1\units    & <350         \\
R32 &                 & NuPure Eliminator 600-CG%
& \hyperref[acr:lu]{LU}     & 1\units    & 150.0+-1.2e3 \\
R33 &                 & \hyperref[acr:saes]{SAES} purifier HP400-903F%
& \hyperref[acr:lu]{LU}     & 1\units    & \sim49e3     \\
        \end{tabular*}%
    \end{ruledtabular}%
\end{table*}%